\documentclass[twocolumn,secnumarabic,amsmath,amssymb, nobibnotes, aps, prb,groupedaddress,superscriptaddress]{revtex4-2}
\usepackage{amsmath,graphicx,latexsym,times,color}
\usepackage{setspace}
\usepackage{hyperref}
\usepackage{array}
\usepackage{textcomp}
\usepackage{titlesec}
\usepackage{physics}
\usepackage{gensymb}
\usepackage{nccmath}
\usepackage{empheq} 
\usepackage{fontawesome5}
\usepackage{enumerate}
\usepackage{textcomp}
\usepackage{gensymb}
\usepackage[cmyk,dvipsnames]{xcolor}
\usepackage[utf8x]{inputenc} 

\usepackage[normalem]{ulem}    
\usepackage[dvipsnames]{xcolor} 

\definecolor{alizarin}{rgb}{0.82, 0.1, 0.26}

\usepackage{lettrine}
\newcommand{\V}[1]{\ensuremath{\mathbf{#1}}} 

\let\oldtimes\times  
\renewcommand\times{{\oldtimes}}

\newcommand\redsout{\bgroup\markoverwith{\textcolor{red}{\rule[0.5ex]{2pt}{0.4pt}}}\ULon}

\usepackage[normalem]{ulem}

\definecolor{darkorchid}{HTML}{bf3eff}

\definecolor{dsgrey}{rgb}{0,1,1}

\newcommand{\remove}[1]{\bgroup\markoverwith{\textcolor{red}{\rule[0.5ex]{2pt}{0.4pt}}}\ULon{#1}\egroup}

\begin{document}

\title{Lifetime of bimerons and antibimerons in two-dimensional magnets}

\affiliation{Universit\'e de Toulouse, CNRS, CEMES, Toulouse, France}   
\affiliation{Institute of Theoretical Physics and Astrophysics, University of Kiel, Leibnizstrasse 15, 24098 Kiel, Germany}  
\affiliation{Science Institute and Faculty of Physical Sciences, University of Iceland, VR-III, 107 Reykjavík, Iceland} 
\affiliation{Kiel Nano, Surface, and Interface Science (KiNSIS), University of Kiel, 24118 Kiel, Germany} 
 
\author{Moritz A. Goerzen}
\affiliation{Universit\'e de Toulouse, CNRS, CEMES, Toulouse, France}
\affiliation{Institute of Theoretical Physics and Astrophysics, University of Kiel, Leibnizstrasse 15, 24098 Kiel, Germany}

\author{Tim Drevelow}
\affiliation{Institute of Theoretical Physics and Astrophysics, University of Kiel, Leibnizstrasse 15, 24098 Kiel, Germany}

\author{Soumyajyoti Haldar}
\affiliation{Institute of Theoretical Physics and Astrophysics, University of Kiel, Leibnizstrasse 15, 24098 Kiel, Germany}
\affiliation{Kiel Nano, Surface, and Interface Science (KiNSIS), University of Kiel, 24118 Kiel, Germany}

\author{Hendrik Schrautzer}
\affiliation{Science Institute and Faculty of Physical Sciences, University of Iceland, VR-III, 107 Reykjavík, Iceland}

\author{Stefan Heinze}
\affiliation{Institute of Theoretical Physics and Astrophysics, University of Kiel, Leibnizstrasse 15, 24098 Kiel, Germany}
\affiliation{Kiel Nano, Surface, and Interface Science (KiNSIS), University of Kiel, 24118 Kiel, Germany}

\author{Dongzhe Li}
\email[Contact author: ]{dongzhe.li@cemes.fr}
\affiliation{Universit\'e de Toulouse, CNRS, CEMES, Toulouse, France}
 
	\date{\today}
	
\begin{abstract}

\noindent Soliton-based computing architectures have recently emerged as a promising avenue to overcome fundamental limitations of conventional information technologies, the von Neumann bottleneck. In this context, magnetic skyrmions have been widely considered for in-situ processing devices due to their mobility and enhanced lifetime in materials with broken inversion symmetry. However, modern applications in non-volatile reservoir or neuromorphic computing raise the additional demand for non-linear inter-soliton interactions. Here, we report that solitons in easy-plane magnets, such as bimerons and antibimerons, show greater versatility and potential for non-linear interactions than skyrmions and antiskyrmions, making them superior candidates for this class of applications. Using first-principles and transition state theory, we predict the coexistence of degenerate bimerons and antibimerons at zero field in a van der Waals heterostructure Fe$_3$GeTe$_2$/Cr$_2$Ge$_2$Te$_6$ -- an experimentally feasible system. We demonstrate that, owing to their distinct structural symmetry, bimerons exhibit fundamentally different behavior from skyrmions and cannot be regarded as their in-plane counterparts, as is often assumed. This distinction leads to unique properties of bimerons and antibimerons, which arise from the unbroken rotational symmetry in easy-plane magnets. These range from anisotropic soliton-soliton interactions to strong entropic effects on their lifetime, driven by the non-local nature of thermal excitations. Our findings reveal a broader richness of solitons in easy-plane magnets and underline their unique potential for spintronic devices.
\emph{}\\
\emph{}

\noindent DOI: $\times \times \times$ \hfill Subject Areas: Condensed Matter Physics, Magnetism, Spintronics
\end{abstract}

\maketitle

\section{INTRODUCTION}

Magnetic skyrmions -- topologically protected quasi-particles with a whirling spin texture in real space -- have raised great attention due to their rich physics and promising applications for future spintronic devices \cite{fert2017magnetic,gobel2021beyond}. Unique properties of skyrmions for applications include their nanoscale size \cite{romming2013writing,meyer2019isolated}, good thermal stability due to their integer topological charge \cite{gobel2021beyond}, efficient manipulation by electric currents \cite{fert2013,sampaio2013nucleation}, all-electrical detection \cite{hanneken2015electrical,
maccariello2018electrical,Perini2019,Dongzhe2023,chen2024all}, and all-optical topological switching \cite{buttner2021observation,dabrowski2022all,khela2023laser}. 
Magnetic bimerons, which emerge as vortex-antivortex pairs in easy-plane magnets, carry the same topological charge
as skyrmions that occur in easy-axis magnets. 
Therefore, bimerons have been interpreted by theory as in-plane counterparts of skyrmions \cite{gobel2019magnetic,Zhang2020a,Shen2020}, and have also been observed experimentally in ultrathin films \cite{jani2021antiferromagnetic,nagase2021observation} and magnetic multilayers \cite{ohara2022reversible}. More recently, bimerons have been predicted in two-dimensional (2D) van der Waals (vdW) magnets in numerous theoretical studies \cite{sun2020controlling,wang2025magnetoelectric,Chen2025}. However, despite the growing interest in bimerons and proposals for spintronic devices
using bimerons \cite{Bo2025}, their thermal stability -- crucial for all applications and quantified by lifetime -- remains largely unexplored.

Calculating the lifetime of magnetic solitons, such as skyrmions and bimerons, is a challenging task that can be addressed using 
transition state theory. In harmonic approximation, one finds the Arrhenius law for the mean lifetime $\tau$ at temperature $T$ \cite{bessarab2012harmonic} 
\begin{equation}\label{eq:Arrhenius}
\tau = \Gamma_0^{-1} \exp\left(\frac{\Delta E}{k_\text{B} T}\right)
\end{equation}
where $\Delta E$ is the energy barrier of the transition and $\Gamma_0$ is the pre-exponential factor which can be explicitly calculated \cite{bessarab2018lifetime}. 
However, many conceptual and technical complications arise due to different possible transition mechanisms, the complexity of identifying the minimum energy 
path in a complex energy surface, and the need to determine not only $\Delta E$ but also $\Gamma_0$ 
which requires identifying and treating Goldstone modes that depend on the type of the considered soliton.
While the collapse and creation of skyrmions has been studied based on various theoretical approaches and by different groups \cite{bessarab2018lifetime,hagemeister2015stability,Desplat2020,Malottki2019,Masell2019,varentcova2020toward,Dongzhe2022_fgt,goerzen2023lifetime,Dongzhe_PRB2024} and even quantitative agreement of skyrmion lifetimes
with experimental data has been demonstrated \cite{muckel2021experimental,goerzen2022atomistic}, the transition mechanisms and lifetime of bimerons have so far not been reported.

Here, we address this gap by presenting the first calculations of the lifetimes of bimerons and antibimerons based on transition-state theory using the prototypical
2D vdW magnet Cr$_2$Ge$_2$Te$_6$
with a honeycomb lattice
as a model system. We obtain the interaction
parameters for the atomistic spin model via
density functional theory (DFT) from
the vdW heterostructure 
Fe$_3$GeTe$_2$/Cr$_2$Ge$_2$Te$_6$ (FGT/CGT), where FGT shows easy out-of-plane axis anisotropy and CGT exhibits an easy-plane anisotropy with exchange frustration.  This
gives rise to the possibility of having multiple topological spin textures in one interface.
In the CGT layer, bimerons and antibimerons emerge. These two topologically distinguishable solitons are symmetry equivalent under a combination of rotation plus spin-inversion and therefore are degenerate at zero field. 

In an external magnetic field, the degeneracy of bimerons and antibimerons is lifted, and we find a homotopical transition to skyrmions and antiskyrmions, respectively. 
We show that the spin structures of bimerons and antibimerons differ in key features from those of skyrmions and antiskyrmions. Quantifying these differences, we identify mechanisms that suggest an increased stability for the easy-plane soliton over those in easy-axis magnets.
The lifetime of solitons, which evolve from bimerons, increases upon applying a magnetic field in contrast to the decrease known from skyrmions. 

Intriguingly, the pre-exponential factor shows almost no change with magnetic field and is identical for bimerons and antibimerons. This is explained by the absence of localized magnetic excitations in the presence of unbroken
$\mathrm{U}(1)$-symmetry, which in our case applies to the rotation of spins within the easy-plane of CGT.
In accordance with the Hohenberg-Mermin-Wagner theorem \cite{mermin1966absence, hohenberg1967existence}, which for 2D materials predicts the existence of gapless magnons in such a situation, we find that the entropy of bimerons and antibimerons, up to a certain field, is dominated by long-range excitations of the magnetization. Thus, only bounded by system dimensions, the pre-exponential factor becomes independent of the exact shape of magnetic solitons in a sufficiently large environment. Based on this fundamental concept, our work contradicts the widespread notion
that bimerons are the 
in-plane counterparts of skyrmions. Using an experimentally feasible vdW interface of FGT/CGT, we 
quantify the effect of the unbroken $\mathrm{U}(1)$-symmetry on the shape and stability of bimerons and antibimerons and further argue that this also increases the degree of non-linearity for soliton-soliton interactions, as required for a wide range of technological applications.

\section{THEORETICAL METHODS}
\label{theory}
\subsection*{A. First-principles atomistic spin model}
To obtain material-dependent magnetic interaction parameters for the vdW interface of FGT/CGT from first-principles, we performed 
DFT calculations using the {\tt FLEUR} code \cite{fleurv26}. We used the generalized Bloch theorem \cite{Kurz2004} and first-order perturbation theory \cite{Heide2009} to calculate exchange energy and Dzyaloshinskii-Moriya interaction (DMI) energy in $\V{q}$ space, respectively. For computational details, see Appendix A. Subsequently, we find that interlayer interactions between FGT and CGT are small and map the DFT total energies of CGT onto the following extended Heisenberg model with an energy function 
\begin{equation}\label{eq:energy_model}
    \begin{split}
        E =  - \sum_{ij}\left[J_{ij}
        (\mathbf{m}_i \cdot \mathbf{m}_j)+\mathbf{D}_{ij}(\mathbf{m}_i \times \mathbf{m}_j)\right] 
        + \sum_{i=1}^N \epsilon_i
        ~. 
    \end{split}
\end{equation}
The spin model includes isotropic Heisenberg exchange 
with constants $J_{ij}$, the DMI with vectors $\mathbf{D}_{ij}\parallel \mathbf{r}_{ij}\times\hat{z}$ and the onsite energies $\epsilon_i$ for each spin are given by
\begin{equation}\label{eq:pot_energy}
    \epsilon_i = -K(m_i^z)^2 -\mu B_zm_i^z ~.
\end{equation}
The onsite energy gives the intrasite energy for each normalized spin $\mathbf{m}_i$ at a discrete lattice site $\mathbf{r}_i$ and contains the magnetocrystalline anisotropy energy (MAE) with constant $K$ as well as Zeeman interaction between an external magnetic field $B_z$ in $z$-direction and an atomic magnetic moment $\mu$.
Dipole-dipole interactions (DDI) are not included in our model. In the CGT monolayer, we find that the DDI is more than an order of magnitude smaller than the MAE, rendering its effect negligible.
All constants for the parameterization of CGT within a honeycomb lattice are found in Table~\ref{table_dmi}. Throughout this work, all simulations are performed in symmetrical simulation boxes with $\sqrt{\frac{N}{2}}\times\sqrt{\frac{N}{2}}$ unit cells, each containing two lattice sites, and periodic boundary conditions.

\begin{figure*}[!tbp]
	\centering
	\includegraphics[width=0.95\linewidth]{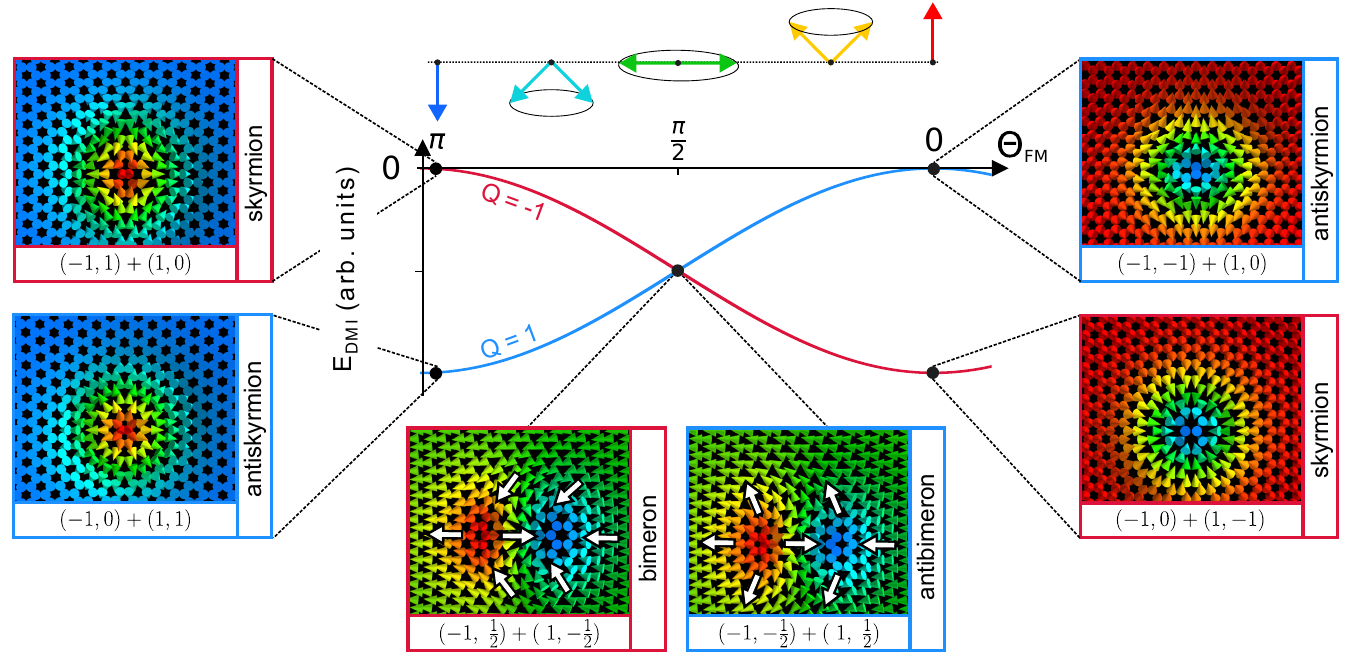}
	\caption{\label{fig:bimeron_skyrmion_transformation} DMI response to (anti-)bimeron and (anti-)skyrmion. Schematic DMI energy $E_{\text{DMI}}$ of solitons with charge $Q=-1$ (red) and $Q=+1$ (blue) under continuous rotation of the ferromagnetic (FM) background by the angle $\Theta_{\text{FM}}$. The orientation of the background for a given $\Theta_{\text{FM}}$ is drawn on top. The circles indicate the space of possible backgrounds that are degenerate because of the unbroken $\mathrm{U}(1)$-symmetry. The respective spin configurations are assigned by arrows and labeled by their meron identifier $\mathrm{idm}(\mathbf{m})$ [cf. Eq.~(\ref{eq:meron_identifier})], which indicates the composition of vorticities $\nu_i$ and polarities $\xi_i$ of the soliton. It is illustrated that bimerons and antibimerons at $\Theta_{\text{FM}}=\pi/2$, which only differ by the polarity of their vortex cores, both exhibit a vortex with $\nu=1$ so that they are degenerate in DMI [cf. Eq.~(\ref{eq:DMI_response1})].
    }
\end{figure*}

\subsection*{B. Transition state theory}
The reaction path for the annihilation of a magnetic soliton into the field-polarized state (e.g., the ferromagnetic (FM) state) is computed using the geodesic nudged elastic band method (GNEB)~\cite{bessarab2015method}. If this method converges, it returns a minimum energy path (MEP) in the space of magnetic configurations that connects a metastable initial state (IN) 
with a final state, which in our case is always the FM state. From the MEP, a first-order saddle point (SP) can be obtained using the climbing image procedure \cite{bessarab2015method} 
\begin{equation}
    \Delta E = E_0^{\text{SP}} - E_0^{\text{IN}} ~.
\end{equation}
For a thermodynamic classification of the stability, not only the IN has to be considered, but also its thermal fluctuations. Therefore, we perform a harmonic approximation of the energy landscape in the vicinity of the IN and the SP 
\begin{equation}\label{eq:harmonic_approx}
    E = E_0 + \frac{1}{2}\sum_{n=1}^{k} \lambda_nq_n^2 + \mathcal{O}(q^3),
\end{equation}
where $q_n$ are normal coordinates. The number $k$ of positive eigenvalues $\lambda_n$
of the Hessian at the IN or the SP is bound by the number of degrees of freedom $k\leq 2N$ for a simulation box with $N$ lattice sites.
This approximation allows for efficient computation of the partition function $Z$ for both states, which are required for the estimation of the free energy $\mathcal{F}$
\begin{equation}
    \mathcal{F} = -k_{\text{B}}T\ln Z~, \quad Z=\int_{E_0}^{\infty} \varrho(E)\mathrm{e}^{-\beta E}~\mathrm{d}E ~.
\end{equation}
Here $\varrho$ is the density of states and $\beta^{-1}=k_{\text{B}}T$.
Note that the partition of states that are degenerate to $E_0$ is computed in zero-mode approximation to the respective degrees of freedom. 
Within the framework of harmonic transition state theory (HTST) \cite{bessarab2012harmonic}, the rate $\Gamma = \Gamma^{\text{IN}\to\text{FM}}$ for the annihilation of a soliton IN can be written in the form of an Eyring equation \cite{goerzen2022atomistic}
\begin{equation}\label{eq:arrhenius_law_entropy}
    \Gamma = \frac{\Lambda^{\ddagger}}{\sqrt{2\pi\beta}}\textrm{e}^{-\beta\Delta \mathcal{F}}~,\quad \Delta \mathcal{F} = \mathcal{F}^{\text{SP}} - \mathcal{F}^{\text{IN}}~.
\end{equation}
The factor $\Lambda^{\ddagger}$ contains the dynamic contributions from the Landau-Lifshitz equation to the rate and reads \cite{potkina2020skyrmions}
\begin{equation}
    \Lambda^{\ddagger} = \frac{\gamma_e}{\mu}\sqrt{\left(\sigma_2\mathbf{v}_{\perp}\right)^T \mathrm{H}_{\text{SP}} \left(\sigma_2\mathbf{v}_{\perp}\right)},
\end{equation}
with $2N\times2N$-Pauli matrix $\sigma_2$, electronic gyromagnetic ratio $\gamma_e$ and Hessian of the SP $\mathrm{H}_{\text{SP}}$ as well as its eigenvector $\mathbf{v}_{\perp}$ associated with the unstable mode so that $\mathrm{H}_{\text{SP}}\mathbf{v}_{\perp} = \lambda_{\perp}\mathbf{v}_{\perp}$ and $\lambda_{\perp}<0$. In this framework, the free energy itself can be expressed in terms of Hessian eigenvalues $\lambda_n$ so that
\begin{equation}
    \Delta \mathcal{F} = \Delta E - \frac{1}{\beta}\ln\left[ \left(\frac{2\pi}{\beta}\right)^{\Delta k/2}\frac{L_{\text{sp}}\prod_{n=1}^{k_{\text{in}}}\sqrt{\lambda_n^{\text{in}}}}{L_{\text{in}}\prod_{n=1}^{k_{\text{sp}}}\sqrt{\lambda_n^{\text{sp}}}}\right]~.
\end{equation}
Linking the free energy $\Delta \mathcal{F} = \Delta \mathcal{U} -T\Delta \mathcal{S}$ to the difference in average thermal energy $\Delta \mathcal{U} = \Delta E + k_{\text{B}}T\Delta k/2$ and entropy $\Delta S$ the rate is most often expressed as the Arrhenius law
\begin{equation}\label{eq:arrhenius_law}
    \Gamma = \frac{\Lambda^{\ddagger}}{\sqrt{2\pi\beta}}\textrm{e}^{\Delta \mathcal{S}/k_{\text{B}} - \Delta k/2}\textrm{e}^{-\beta\Delta E} = \Gamma_0\textrm{e}^{-\beta\Delta E}~.
\end{equation}
The form of the internal energy used above follows directly from the equipartition theorem for any system with energy as given in Eq.~(\ref{eq:harmonic_approx}).

\section{RESULTS}

\subsection*{A. Bimerons and antibimerons in 2D magnets}

In order to point out the advantage of easy-plane magnets over easy-axis magnets in terms of hosting degenerate magnetic particles, we study not only the stability of bimerons but also antibimerons. These states are magnetic configurations $\mathbf{m}:\mathbb{R}^2\to\mathbb{S}^2$ which exhibit 
opposite topological charges $Q=\pm1$. In accordance with Refs.~\cite{nagaosa2013topological,kuchkin2020turning}, we define particles and antiparticles in analogy to electrons and positrons, so that bimerons and skyrmions possess a charge of $Q=-1$ and antibimerons and antiskyrmions possess $Q=1$ as illustrated in Fig.~\ref{fig:bimeron_skyrmion_transformation}. Note that this does not coincide with the definition in other works \cite{jiang2017direct}. The difference between bimerons and antibimerons becomes apparent when imagining them as bound states of two merons, as depicted in Fig.~\ref{fig:bimeron_skyrmion_transformation}. Each meron is characterized by a vortex- or winding number $\nu_i\in\mathbb{Z}$ and the polarity of its core $\xi_i=\frac{1}{2}\left[\cos\Theta(\rho)\right]^{\infty}_0$ against the polarized background at a radial distance $\rho=\infty$ (see Appendix A). In this picture, we can write the meron-resolved topological charge
\begin{equation}\label{eq:topological_charge1}
    Q(\mathbf{m}) = \frac{1}{4\pi}\int_{\mathbb{R}^2} \mathbf{m}\left(\frac{\partial\mathbf{m}}{\partial x}\times\frac{\partial\mathbf{m}}{\partial y}\right)~\mathrm{d}^2\mathbf{r} = \sum_{i=1,2}\nu_i\xi_i~.
\end{equation}
The meron-resolved decomposition is well supported considering the the topological invariant $\Tilde{Q}_i=(\nu_i^{\text{t}}, \nu_i^{\text{b}})$ of a single meron in the second homotopical group $\pi_2(\mathbb{S}^2,\mathbb{S}^2\setminus \{P_1,P_2\})=\mathbb{Z}\times\mathbb{Z}$ with $P_i=\pm\hat{\mathbf{z}}$ \cite{rybakov2025topological, zhu2026light}. Practically $\nu_i^{\text{t}}$ is the winding around a center with polarization $P_1=\hat{\mathbf{z}}$ and $\nu_i^{\text{b}}$ from a center with $P_2=-\hat{\mathbf{z}}$. In this way, the topological invariant for the bound state of two merons, namely bimerons and antibimerons, is decomposed as
\begin{equation}\label{eq:meron_identifier}
    \Tilde{Q} = (\nu_1^{\text{t}}, \nu_1^{\text{b}}) + (\nu_2^{\text{t}}, \nu_2^{\text{b}})~,\quad Q = \frac{1}{2}\sum_{i=1}^2 \nu_i^{\text{t}} - \nu_i^{\text{b}}~.
\end{equation}
For the purpose of describing the transformation between (anti-)bimerons and (anti-)skyrmions, which additionally requires information on the canted background, we make use of the fact that the $Q$ in Eq.~(\ref{eq:meron_identifier}) and Eq.~(\ref{eq:topological_charge1}) take the same value for identical textures. Thus, we choose the representation $(\nu_i^{\text{t}}, \nu_i^{\text{b}})\leftrightarrow(\nu_i, \xi_i)$ which takes into account the equivalence $(\pm1, 0)\leftrightarrow(\mp1, \frac{1}{2})$ and $(0, \pm1)\leftrightarrow(\mp1, -\frac{1}{2})$. In this representation $\xi$ indicates the difference between a (anti-)meron with $\xi=\pm\frac{1}{2}$ and a (anti-)skyrmion with $\xi=\pm1$. Note that for bound states of two merons, the restriction to vortex-antivortex pairs with $\nu_1=-\nu_2$, which is inevitable for a constant magnetization at $\rho=\infty$ \cite{kuchkin2020turning} and thus also for periodic boundary conditions at the edge of our simulation boxes, requires $\xi_1-\xi_2=\pm1$ in order to return integer values of $Q$. As indicated by the white arrows in the sketches of Fig.~\ref{fig:bimeron_skyrmion_transformation}, the difference between bimerons with $Q=-1$ and antibimerons with $Q=-1$ arises from the composition of $\xi_i$ and $\nu_i$, giving rise to distinct topological charges according to Eq.~(\ref{eq:topological_charge1}).

\begin{figure*}[t]
	\centering
	\includegraphics[width=1.0\linewidth]{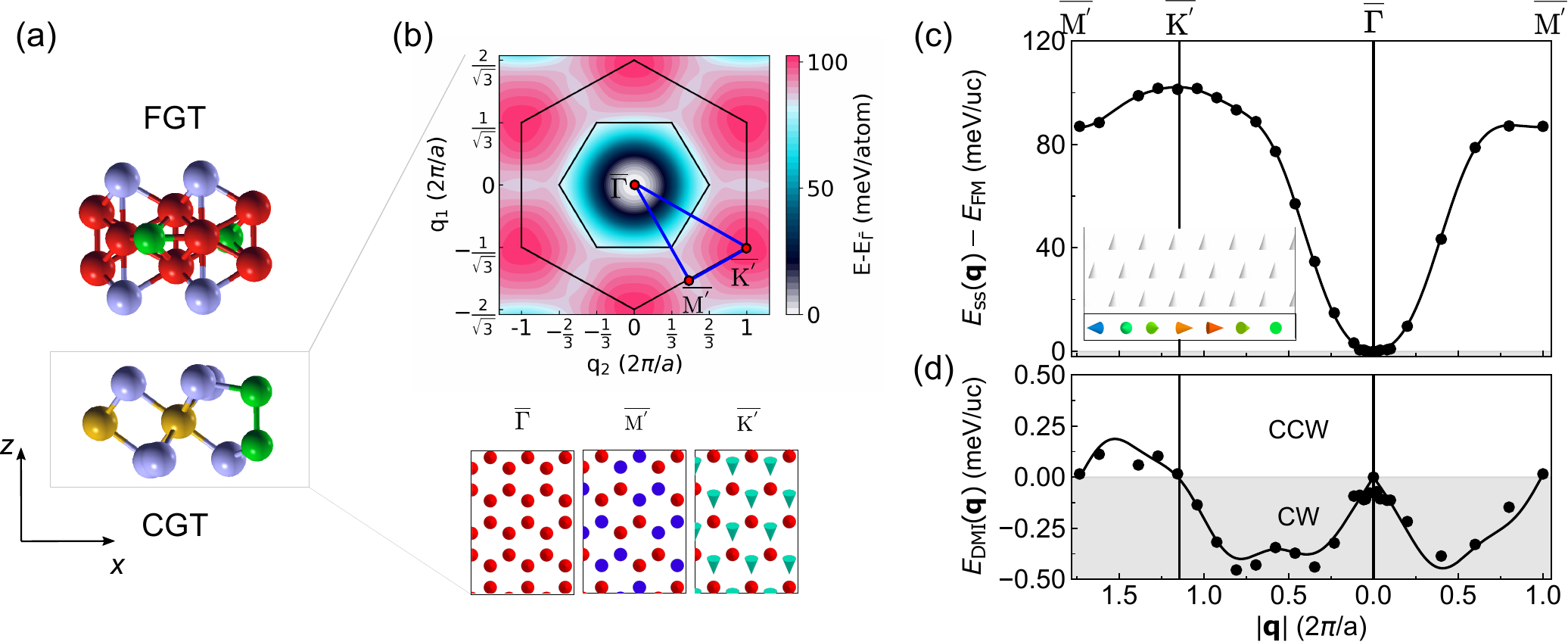}
	\caption{\label{spin-spiral} DFT total energy calculations for spin spirals in the FGT/CGT heterostructure. (a) Side view of the atomic structure of the FGT/CGT
    interface. (b) Exchange energy with respect to the $\overline{\Gamma}$-point 
    (FM state) in the 2D Brillouin zone for the
    honeycomb lattice model of the CGT layer with the two high symmetry paths $\overline{\Gamma \text{M}^{'}}$ and $\overline{\Gamma \text{K}^{'}\text{M}^{'}}$ indicated. 
    Lower panels show the spin structures in the CGT layer obtained at the $\overline{\Gamma}$-point, the $\overline{\rm M'}$-point (RW-AFM state),
    and the $\overline{\rm K'}$-point
    ($120^\circ$ N\'eel state).
    (c) Spin spiral energy dispersions
    without SOC, $E_{\text{SS}}(\V{q})$ and 
    (d) SOC-induced DMI energy, $E_{\text{DMI}}(\V{q})$  for the CGT layer. Black circles correspond to the DFT total energies, while black curves are the fits to the extended Heisenberg model up to the seventh NN. All energies are given with respect to the FM state at $\V{q} = 0$. Note that positive and negative DMI energy contributions represent 
    clockwise
    (CW) and 
    counter-clockwise
    (CCW) cycloidal spin spiral configurations. 
    Inset in panel (c) shows 
    flat spin spirals in the CGT layer with the FGT layer fixed in a ferromagnetic state, enabling extraction of exchange and DMI in CGT.
    The top three layers represent the Fe atoms in the FGT layer, while the bottom layer represents the Cr atoms in the CGT layer.
    }
\end{figure*}
\begin{table*}[htb]
	\centering
	\scalebox{0.95}{
		\begin{tabular}{ccccccccccccccccccc}
			\hline\hline
layer & lattice & $J_1$ & $J_2$ & $J_3$ & $J_4$ & $J_5$ & $J_6$ & $J_7$ & $J_8$ & $D_1$ & $D_2$ & $D_3$ & $D_4$ & $D_5$ & $D_6$ & $D_7$ & $K$ & $\mu$\\ 
\hline
CGT& honeycomb & 20.45 & 1.88 & 4.0 & $-$0.45 & $-$0.35 & $-$0.13 & $-$0.48 & 0.1 & 0.249 & $-$0.013 & 0.007 & 0.025 & $-$0.01 & 0.012 & $-$0.019 & $-$0.13 & 3.06 \\
			\hline
	\end{tabular}}
    \caption{Magnetic interaction constants obtained from DFT for the CGT layer of the FGT/CGT heterostructure. Shell-resolved Heisenberg exchange constants ($J_i$) and DMI constants ($D_i$) obtained by fitting the energy contribution to spin spirals without and with SOC from DFT calculations as presented in Fig.~\ref{spin-spiral}. $J_i>0$ ($J_i<0$) corresponds to FM (AFM) exchange, $D_i>0$ ($D_i<0$) favors CW (CCW) cycloidal spirals, and $K<0$ stabilizes in-plane magnetization. The interaction constants on a honeycomb lattice can be directly calculated using the given weighting factors (cf.~Appendix A). All values are given in meV/atom, except for the spin moment $\mu$ of Cr, which is expressed in $\mu_{\text{B}}$/atom.
    }  \label{table_dmi} 
\end{table*}	

\begin{figure*}[!tbp]
	\centering
	\includegraphics[width=1.0\linewidth]{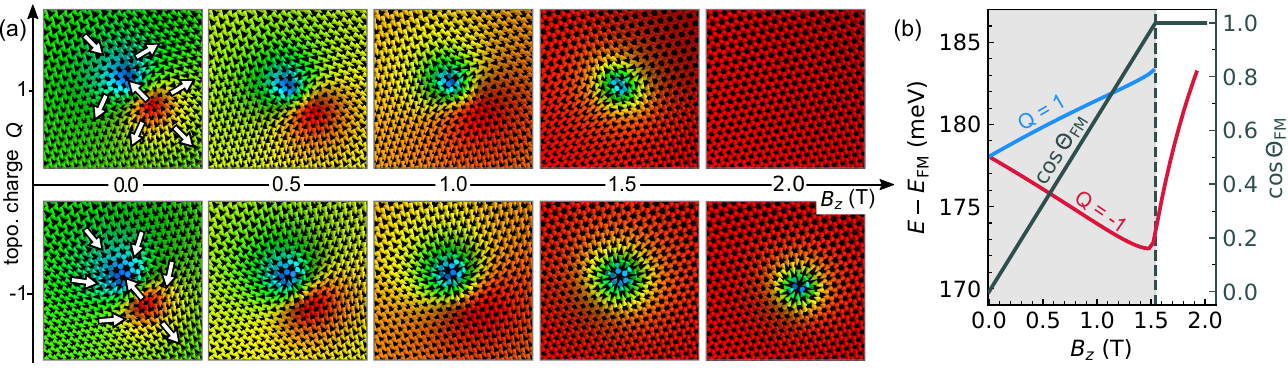}
	\caption{\label{bimeron_skyrmion_transformation} (a) (Anti-)bimeron to (anti-)skyrmion transformation in an all-magnetic FGT/CGT vdW interface. Magnetic configuration of metastable solitons with topological charges $Q=+1$ (upper row) and $Q=-1$ (lower row) starting from antibimeron ($Q=+1$) and bimeron ($Q=-1$) at $B_z=0.0~$T up to skyrmion-/antiskyrmion-like configurations at $B_z\geq1.5~$.
    The difference between bimeron and antibimeron is indicated by the arrows, which schematically show the in-plane component of the magnetization at characteristic points on the soliton circumference. Independent of $Q$, an applied magnetic field $B_z>0~$T induces a canting of the background. In this way, merons with positive polarity vanish into the background, leaving only merons with negative polarity $\xi$. Since these configurations have either vortex ($\nu=1$) or antivortex ($\nu=-1$) characters, they respond differently to DMI (cf. Fig. \ref{fig:bimeron_skyrmion_transformation}). 
    (b) The energy of the spin textures of $Q=-1$ (red) and $Q=+1$ (blue) shown in (a) is displayed as a function of the magnetic field $B_z$. 
The offset $E_{\text{FM}}$ is the energy of the polarized state for the respective magnetic field, 
whose $m_z$-component (black), $\cos\Theta_{\text{FM}}$, is displayed on the right axis.
    }
\end{figure*}

It is important to note that all interactions in the energy function in Eq.~(\ref{eq:energy_model}), 
apart from Zeeman interaction and the DMI, are invariant under spin inversion $\mathbf{m}\to-\mathbf{m}$. Since Eq.~(\ref{eq:topological_charge1}) and (\ref{eq:meron_identifier}) state that the inverted state $\Tilde{Q}(-\mathbf{m})=(\nu_1,-\xi_1) + (\nu_2, -\xi_2)$ has to have inverted topological charge, it follows that bimerons and antibimerons at zero field are degenerate under exchange and anisotropy. The invariance under DMI requires an additional change in the helicity of the merons, which is subject to the next section.

\subsection*{B. Distinction from (anti-)skyrmions}

The degeneracy of bimerons and antibimerons is in strong contrast to skyrmions and antiskyrmions.
The major difference between (anti-)bimeron and (anti-)skyrmion is their response to DMI. This can be seen from the continuum formulation for interfacial DMI of a perfectly radial symmetric (anti-)skyrmion
\begin{equation}\label{eq:DMI_response1}
    E_{\text{DMI}}=2 \pi \mathcal{D} \delta_{\nu, 1} \cos \gamma \int_{0}^{\infty}\left(\frac{\partial \Theta}{\partial \rho} + \frac{\sin \Theta \cos \Theta}{\rho}\right)  \rho \mathrm{d} \rho ~,
\end{equation}
which shows a dependency on the helicity $\gamma$ and the vortex number $\nu$ via the Kronecker-Delta term $\delta_{\nu, 1}$ (see Appendix A.3 for details on soliton profile). 
In this idealized case, the DMI with a magnitude given by $\mathcal{D}$ in the continuum model, therefore, only affects the energy of solitons with vortex number $\nu=1$. This can account for (anti-)skyrmions but also for isolated (anti-)merons. 
Bimerons and antibimerons, on the other hand, both possess a meron with a favored vortex number due to the vortex-antivortex restriction to Eq.~(\ref{eq:meron_identifier}). The evolution of the DMI under continuous transformation between (anti-)bimeron and (anti-)skyrmion is schematically illustrated in Fig.~\ref{fig:bimeron_skyrmion_transformation}. It is shown, that bimerons and antibimerons are stabilized equally by DMI, while a homogeneous rotation of all spins by $\Theta_{\text{FM}}$ around the $y$-axis lifts this degeneracy until the maximum difference is reached for skyrmions with favored remaining composition $\Tilde{Q}(\mathbf{m})=(1,-1)$ and antiskyrmions with a composition $\Tilde{Q}(\mathbf{m})=(-1,-1)$ that is ignored by the DMI in Eq.~(\ref{eq:DMI_response1}). Notably, the two states with $\nu=1$ in Fig.~(\ref{fig:bimeron_skyrmion_transformation}) (upper left and lower right) differ in their helicity by $\Delta\gamma=\pi$. This reflects that the inversion $\mathbf{m}\to-\mathbf{m}$ requires an additional change in helicity in order to obtain degenerated soliton states, as mentioned in the previous section.

\subsection*{C. Spin spiral dispersions}

Having understood the fundamental difference between (anti-)skyrmions and (anti-)bimerons discussed in the previous subsection, let us move to a realistic first-principles-based spin model. In order to obtain material-dependent
interaction constants, our chosen system is an all-magnetic vdW heterostructure composed of FGT and CGT layers, which was synthesized experimentally recently \cite{wu2022van}, as shown in Fig.~\ref{spin-spiral}(a). Here, we calculate all the required parameters in Eq.~(\ref{eq:energy_model}) based on DFT calculations. Since the hybridization between FGT and CGT is rather small due to their vdW gap nature, we treat them in our spin model as separate layers and neglect the interlayer coupling. The validity of this approximation is justified in Appendix B through explicit calculations of interlayer exchange interactions. On the other hand, it is worth noting that our model includes the hybridization effect at the interface between FGT and CGT. For the scope of this work, we only model the honeycomb lattice of the Cr atoms in CGT, which, due to easy-plane MAE with $K<0$ and frustrated exchange, serves as a suitable platform for hosting bimerons and antibimerons.

We plot in Fig.~\ref{spin-spiral}(b) the exchange energy map in $\V{q}$ space  
as calculated based on the exchange constants obtained via DFT. Three representative spin spiral states are shown for high symmetry points of the 2D Brillouin zone (BZ): the FM state corresponding to the $\overline{\Gamma}$-point, the row-wise antiferromagnetic state at the $\overline{\text{M}^{'}}$-point, and the $120^\circ$ Néel state at the $\overline{\text{K}^{'}}$-point. Fig.~\ref{spin-spiral}(c) displays the spin spiral curve without spin-orbit coupling (SOC) obtained via DFT along the high symmetry paths of the 2D BZ for the CGT layer, i.e., we fix all Fe spins to the FM state and consider a spin spiral only in the CGT layer with a rotational plane perpendicular to the Fe spins [see inset of Fig.~\ref{spin-spiral}(c)]. Here, $E(\V{q})$ is calculated using the generalized Bloch theorem \cite{Kurz2004}. Note that an extended $\V{q}$ zone beyond the first BZ is necessary to accurately fit the DFT total energies because of the honeycomb lattice of CGT (see Appendix A).   We find that the FM state is energetically lowest and that the spin moments are almost constant with respect to $\V{q}$ (not shown), indicating the applicability of our spin model for the CGT layer. 

By including SOC in our DFT calculations within first-order perturbation theory, we can determine the energy contribution of DMI to the spin spiral dispersion \cite{Heide2009}. In the CGT layer, the DMI favors cycloidal spin spirals with a clockwise (CW) rotational sense, i.e.~$E_{\text{DMI}}(\V{q})<0$ [Fig.~\ref{spin-spiral}(d)]. The corresponding micromagnetic DMI amplitude of the CGT layer is approximately 0.10 mJ/m$^2$.
The magnetic ground state remains ferromagnetic (FM) when both the DMI and MAE are taken into account, i.e., $E_{\text{ss}}(\mathbf{q}) + E_{\text{DMI}}(\mathbf{q}) + |K|/2$, where $K = -0.136$ meV/atom (in-plane). By fitting our DFT data (filled dots) to the spin model (curves) in Eq.~(\ref{eq:energy_model}), we obtain magnetic interaction parameters for arbitrary NN as listed in Table \ref{table_dmi}. In the following, all atomistic spin simulations are performed using the magnetic interactions from Table \ref{table_dmi} on the honeycomb lattice to model solitons in the CGT layer.

\subsection*{D. Field-induced transformation}

Interestingly, as predicted in Fig.~\ref{fig:bimeron_skyrmion_transformation}, our atomistic spin simulations demonstrate the emergence of bimeron–skyrmion and antibimeron–antiskyrmion transformations under an external magnetic field in the CGT layer, as shown in Fig.~\ref{bimeron_skyrmion_transformation}(a) for soliton configurations at selected magnetic fields. The external magnetic field induces a canting of the magnetic background, such that for high fields the (anti-)bimerons transform into (anti-)skyrmions. In such a background, skyrmion-like textures are energetically favored due to their vorticity $\nu=1$, as illustrated in Fig.~\ref{bimeron_skyrmion_transformation}(b).

Such a rotation of the magnetic background arises from the competition between Zeeman interaction and the MAE \cite{yu2024spontaneous}, and the associated canting angles can be estimated. This competition is quantified in the potential energy given in Eq.~(\ref{eq:pot_energy}).
A homogeneous magnetization with $z$-component $m_z=\cos\Theta_{\text{FM}}$ that minimizes the potential energy while being subject to the spherical constraint $m_z^2\leq1$ can be obtained as the stationary point of the accordingly constructed Lagrangian and reads 
\begin{equation}\label{eq:background_mz}
    \cos\Theta_{\text{FM}} = -\mu B_z/2K.
\end{equation}
Due to the constraint, this linear canting of the FM background with magnetic field strength $B_z$ is only valid up to a saturation field of $B_z^{\text{sat}}=-2K/\mu \approx 1.54~\text{T}$ in CGT with $\mu=3.06~\mu_{\text{B}}$/atom and $K=0.136$~meV/atom. This behavior of the magnetic background is numerically validated and displayed in Fig.~\ref{bimeron_skyrmion_transformation}(b). Further we obtain metastable soliton solutions with $Q=-1$ (bimeron $\to$ skyrmion)
up to $B_z\approx1.92~$T, far beyond $B_z^{\text{sat}}$, and solitons with $Q=+1$ (antibimeron $\to$ antiskyrmion) up to $B_z\approx1.52~$T. This leads to a smooth transformation between isolated (anti-)bimerons and (anti-)skyrmions, controlled by $B_z$. Beyond $B_z^{\text{sat}}$, the FM background is polarized perpendicular to the easy plane. Finally, as expected, increasing the magnetic field strength beyond these limits leads to the annihilation of the respective solitons.

\begin{figure*}[tp]
	\centering
	\includegraphics[width=0.95\linewidth]{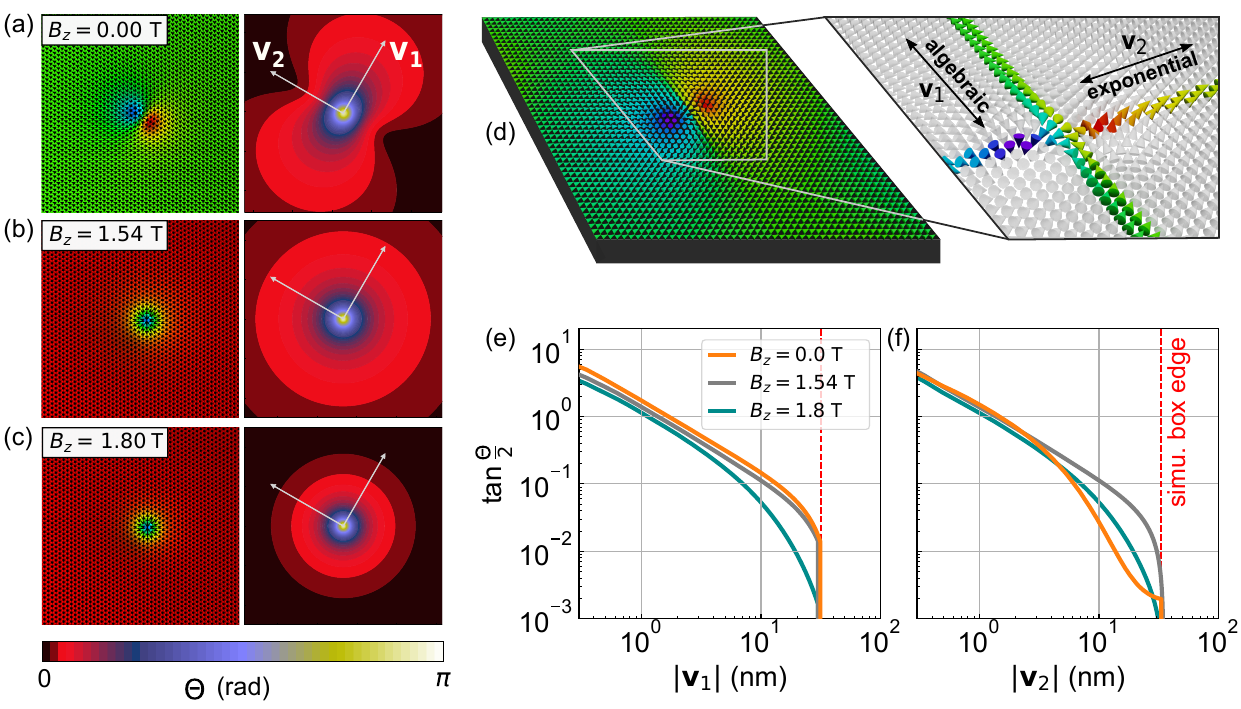}
	\caption{\label{fig:profiles} Soliton $\Theta$-profiles in a FM background
    with canted magnetization.
    (a) Equally sized images of a metastable bimeron at $B_z=0~$T and its $\Theta$-profile. The $\Theta$-profiles are given with respect to the canted FM background. Similarly, the magnetization and profile for $B_z=1.54~$T (corresponds to the saturation field $B_z^\text{sat}$) are shown in (b) and for $B_z=1.8~$T in (c). (d) 
    Bimeron at $B_z=0~$T with zoom on its core region. We highlight magnetization rotations along $\mathbf{v}_1$ and $\mathbf{v}_2$, illustrating that the magnetization cross-sections can be approximated as domain walls rotating $xz$- and the $xy$-plane, respectively. 
    The $\Theta$-profiles of solitons from (a), (b), (c) along the direction $\mathbf{v}_1$ are shown in (e) and along $\mathbf{v}_2$ in (f). All profiles are displayed over a double logarithmic axis so that an algebraic decay $\propto~ 1/\rho$ becomes apparent by a line with a slope of $-1$. The vertical red line indicates where the $100\times100$ simulation box ends. It shall be illustrated that the soliton at $B_z=1.54~$T (grey) decays algebraically in both directions, while the skyrmion-like state at $B_z=1.8~$T (green) decays faster, approximately exponential. In contrast, the bimeron at $B_z=0~$T (orange) decays algebraically along $\mathbf{v}_1$ and exponentially along $\mathbf{v}_2$. 
    }
\end{figure*}


\subsection*{E. Localization in easy-plane magnets}\label{sec:localization}

Regarding the usually strong correlation between soliton size and lifetime, it is mandatory to point out qualitative differences between soliton shapes in different regimes of $B_z$. For that purpose, we define the $\Theta$-profile with respect to the orientation of the FM background
\begin{equation}\label{eq:relative_theta}
    \Theta(\mathbf{r}) = \arccos\left[\mathbf{m}(\mathbf{r})\cdot\mathbf{m}_{\text{FM}}\right]~, 
\end{equation}
which is thus applicable in arbitrarily canted backgrounds. Figs.~\ref{fig:profiles}(a)-(c) show $\Theta$-profiles of solitons with $Q=-1$ at three exemplary magnetic field strengths chosen, including $B_z^{\text{sat}}$, which is also known as ``point of critical
coupling" \cite{schroers2019gauged, barton2020magnetic, kuchkin2020magnetic} where the onsite energy from Eq.~(\ref{eq:pot_energy}) for each spin becomes
\begin{equation}\label{eq:bogomolnyi_fingerprint}
    \epsilon_i \big|_{B_z = B_z^{\text{sat}}} 
    = K - K\,(1-m_i^z)^2 ,
\end{equation}
with the saturation field defined by $B_z^{\text{sat}} = -2K/\mu$.

In this situation, it can analytically be shown that the $\Theta$-profile of skyrmions in the $+m_z$-polarized background decays algebraically \cite{barton2020magnetic} instead of by an exponential double domain wall profile 
\begin{equation}\label{eq:theta_profile}
    \Theta(\mathbf{r}) = \left\{ \begin{array}{ll}
        2\arctan\left(\frac{\kappa}{\rho}\right) &  \text{algebraic}\\
        2\sum_{\pm} \arctan\left(\mathrm{e}^{\frac{-\rho\pm c}{w}}\right) & \text{exponential} 
    \end{array}\right.,
\end{equation}
with constants $\kappa, c, w$.
This is reflected by the spin textures in Fig.~\ref{fig:profiles}(b) and (c), which
look similar at first glance. However, their $\Theta$-profiles reveal significant differences. 

In order to quantify the differences, the profiles are evaluated
along the two perpendicular directions 
$\mathbf{v}_1$ and $\mathbf{v}_2$, depicted by white arrows and the results are displayed in Fig.~\ref{fig:profiles}(e),(f).
By displaying the profile on a double logarithmic axis, an algebraic decay $\propto 1/\rho$ is visible as a straight line of slope $-1$. This is found for the skyrmion at $B_z^{\text{sat}}$ (grey) 
until reaching the boundary of the simulation box, which in turn is indicated by a vertical red line. For the state at $B_z=1.8~$T (green), we find that it decays faster than $1/\rho$, approximately exponentially. 
However, the most interesting case is the bimeron at $B_z=0~$T. It can be seen that the profile converges faster than $1/\rho$ along $\mathbf{v}_2$ [cf. Fig.~\ref{fig:profiles}(f)] but algebraically along $\mathbf{v}_1$ [cf. Fig.~\ref{fig:profiles}(e)].

Despite the lack of an analytical solution for the $\Theta$-profile at $B_z=0~$T this behavior can be understood when considering that the magnetization of the bimeron along $\mathbf{v}_1$ and $\mathbf{v}_2$ can be modeled as double domain walls, illustrated in Fig.~\ref{fig:profiles}(d).
Because of the in-plane MAE, the domain wall along $\mathbf{v}_2$, which rotates in the plane spanned by $\mathbf{v}_2$ and $\hat{\mathbf{z}}$, feels an effective easy-axis anisotropy $K_2=-K$ and therefore exhibits a typical exponential domain wall profile \cite{rohart2013skyrmion} with finite width $\sqrt{J /K_2}$. The domain wall along $\mathbf{v}_1$, on the other hand, which describes the boundary between the two merons, rotates within the easy plane and does not feel any effective anisotropy.
The domain walls' unique dependence on exchange then enforces an algebraic decay, similar to that of Belavin-Polyakov solitons \cite{polyakov22metastable}.

We conclude that solitons at fields $B_z\leq B_z ^{\text{sat}}$ decay more slowly along at least one direction, in a way that sufficient convergence of the energy is found for sample sizes of $\sim5~$mm$^2$, which is far beyond reasonable numerical effort within either the atomistic or micromagnetic spin simulation framework. 
However, since simulations are affected by finite-size effects, the convergence of the DMI energy in Eq.~(\ref{eq:DMI_response1}) for both profiles in Eq.~(\ref{eq:theta_profile}) indicates that the solitons are localized and that reliable numerical results can be obtained when extrapolated to infinite simulation boxes. Overall, we conclude that skyrmions are strongly isotropically localized, while bimerons exhibit anisotropic nonlocality until they reach convergence in an extremely large supercell (not calculable in practice).

\begin{figure*}[tbp]
	\centering
	\includegraphics[width=1\linewidth]{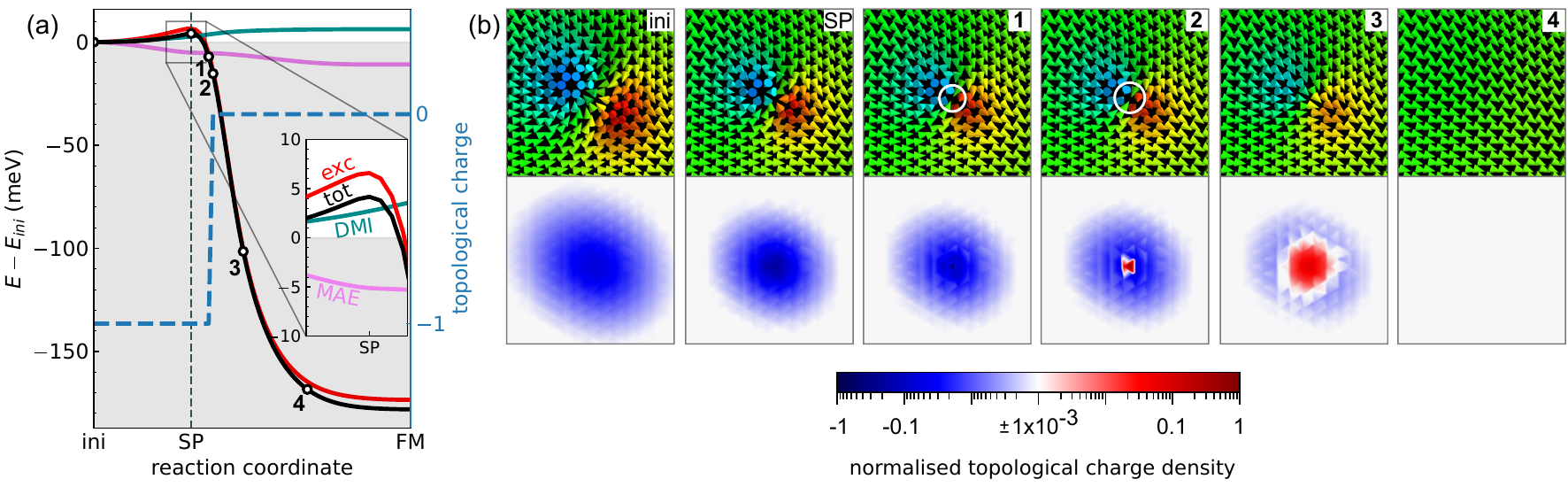}
	\caption{\label{fig:bimeron_mep} Minimum energy path and transition mechanism of bimeron decay. (a) Energy of images along the minimum energy path for the transition between bimeron and FM (black) at $B_z = 0$~T. The energy is decomposed into exchange (red), DMI (green), and MAE (violet). The topological charge (blue dashed) of images is plotted with respect to the right axis. (b) Selected spin textures (top panels) along the MEP. The numbering denotes the position of images as indicated in (a). The corresponding topological charge densities are also shown (lower panels). Due to the non-uniform triangulation of the honeycomb lattice, the charge is normalized with respect to the triangle area, as described in Appendix A. The sign switching of the topological charge density in the center, between images 1 and 2, marks the disappearance of the total topological charge. The saddle point (SP) and Bloch point (BP) do not coincide. The bimeron collapse mechanism at $B_z=0~$T deviates from that of a skyrmion only by a slight elliptical deformation.
    }
\end{figure*}

\subsection*{F. Bimeron collapse mechanism}

To obtain a meaningful measure for the thermal stability of solitons in CGT, we compute their average lifetime [cf. Eq.~(\ref{eq:Arrhenius})] within the framework of HTST \cite{bessarab2012harmonic}. The general procedure is described in Section \ref{theory} and requires knowledge about the collapse mechanism for the annihilation of a soliton. An example of the energy along a MEP for the annihilation of a bimeron with $Q=-1$ at $B_z=0$~T is shown in Fig.~\ref{fig:bimeron_mep}(a). 
From the decomposition of the energy along the path into individual interactions, it becomes apparent that the energy barrier is dominated by exchange (red), arising from frustration between short- and long-range interactions as visible in Table \ref{table_dmi}: short-range FM coupling ($J_1, J_2, J_3$) competes with long-range AFM coupling ($J_4, J_5, J_6, J_7$).

Contributions from DMI (cyan) and MAE (violet)
play a minor role in the stability of bimerons. 
In the collection of images along the MEP shown in Fig.~\ref{fig:bimeron_mep}(b), the state containing the Bloch point (BP) is between the images labeled 1 and 2, but not sampled directly. This position along the MEP is characterized as the point where the total topological charge switches, in our case, accompanied by a local change of sign in the topological charge density (see Appendix A for computational details), which is visible in the topological density map below the corresponding configurations [Fig.~\ref{fig:bimeron_mep}(b)].
The SP, in contrast, is the image of the highest total energy and seems to match the image, where the minimum distance between the two meron centers is evidently reached. We also observe that the SP is different from the BP. 
Moreover, the two-meron picture for the collapse is also supported by the elliptic shape of the topological density along the path, which distinguishes it from the radial collapse known for skyrmions \cite{Malottki2019, muckel2021experimental}.
Having understood the anisotropic nonlocality of bimerons and their collapse mechanism, in the following subsection, we explain how to define the size of bimerons, going beyond standard skyrmions widely studied in the literature. We will also carefully address how these large finite-size effects influence key properties of bimerons, such as their stability, the dynamical contribution, and the entropy difference.

\begin{figure*}[htp]
    \centering
    \includegraphics[scale=0.46]{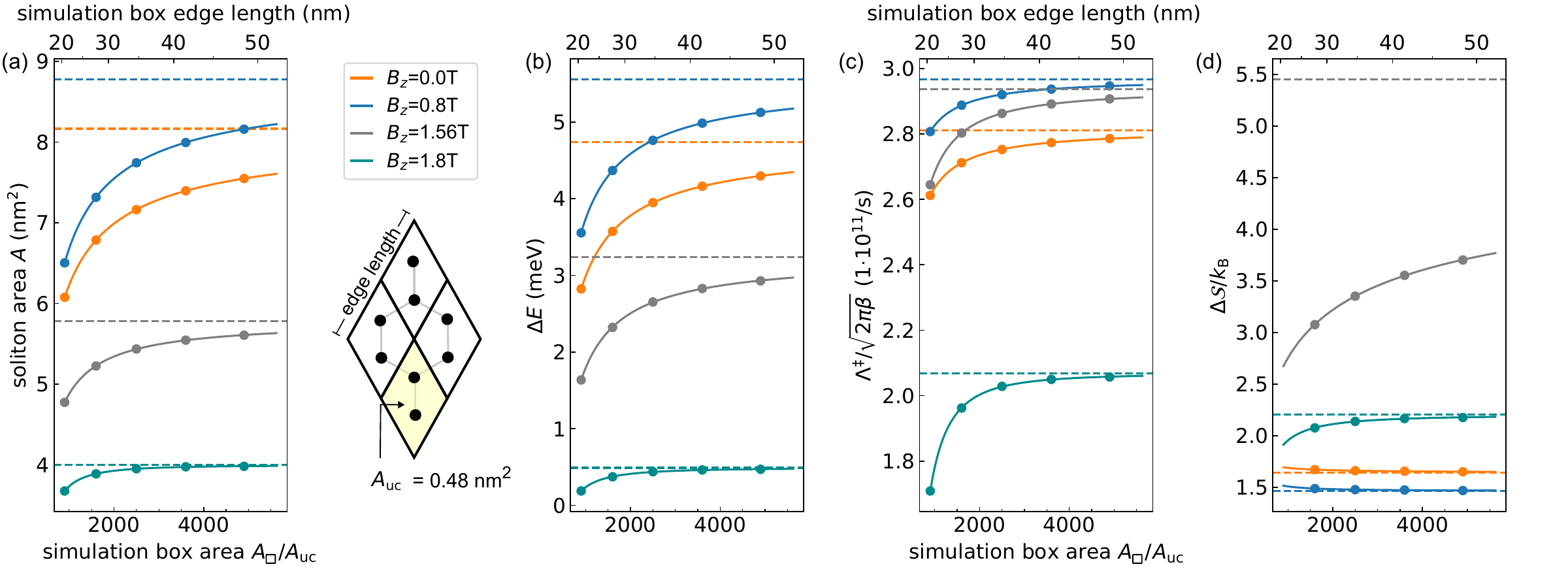}
    \caption{Magnetic-field-dependent spatial convergency of solitons. (a) Area of solitons $A$ at specific magnetic fields for different simulation box sizes $A_{\Box}$ in units of the unit cell area $A_{\text{uc}}$, which, for the honeycomb lattice construction used in this work, is sketched on the right of panel (a), $A_\text{uc} = 0.48~\text{nm}^2$. Four magnetic fields are selected: $B_z = 0.0~\text{T}$ (the bimeron state), $B_z = 0.8~\text{T}$ (the intermediate state between zero field and saturation field), $B_z = 1.56~\text{T}$ (near the saturation field), and $B_z = 1.8~\text{T}$ (the skyrmion state). The lines are obtained by fitting Eq.~(\ref{eq:fit}) against numerical data, and the horizontal lines represent the obtained limit $f_{\infty}$. Similar to the area, we display the convergence of the energy barrier $\Delta E$ (b), the dynamical factor $\Lambda^{\ddagger}/\sqrt{2\pi\beta}$ (c), and the dimensionless entropy difference $\Delta S/k_{\text{B}}$ (d). Due to the intrinsic temperature dependence, the dynamical and entropic contributions are shown for $T = 1~\text{K}$.  
    }
    \label{fig:convergency}
\end{figure*}

\subsection*{G. Convergence of algebraically decaying solitons}
\label{finite-size-effect}

An often used procedure for the estimation of the radius $R$ of radially symmetric solitons, especially skyrmions \cite{bocdanov1994properties}, is based on the inflection point $\rho_c$ of the $\Theta$-profile, where $\ddot{\Theta}(\rho_c)=0$ and
\begin{equation}\label{eq:skyrmion_radius}
    R= \rho_c - \frac{\Theta(\rho_c)}{\dot{\Theta}(\rho_c)} ~.
\end{equation}
Due to the broken radial symmetry of solitons at $B_z\leq B_z ^{\text{sat}}$ and the fact that algebraic profiles [cf. Eq.~(\ref{eq:theta_profile})] do not obey an inflection point, an invariant approach is needed. Because of this, we give an estimate of all soliton sizes not in terms of a radius but rather of the soliton area $A$. This is in contrast to previously used methods for the size estimation of bimerons \cite{kharkov2017bound}. 

We interpolate the $\Theta$-profile [cf. Eq.~(\ref{eq:theta_profile})] of metastable solitons by cubic polynomials to obtain a continuous function $\Theta(\mathbf{r})$. This function is then sampled by a fixed number of $X_{\Box} = 4\cdot10^6$ 
discrete points on a square superlattice that superimpose the simulation box with area $A_{\Box}$. The area of the soliton is then computed as the ratio of the number $X(A_{\Box})$ of sampling points within the contour $\Theta(\mathbf{r})\geq\pi/2$ and the total number of sampling points $X_{\Box}$.
Due to the slow spatial convergence of the solitons and accompanying finite-size effects, this estimation is performed for different system sizes and extrapolated to infinitely large simulation boxes so that the final procedure comes down to
\begin{equation}\label{eq:area_estimation}
    A = \int_{\mathbb{R}^2} H\left[\Theta(\mathbf{r})-\frac{\pi}{2}\right]~\mathrm{d}^2\mathbf{r} \approx \lim_{A_{\Box}\to\infty} A_{\Box}\frac{X(A_{\Box})}{X_{\Box}}~,
\end{equation}
with the Heavyside function $H$.
Thereby, because of $\Theta$ being defined with respect to the polarized background [cf. Eq.~(\ref{eq:relative_theta})] $A$ is invariant under the global rotation of the magnetization.  In addition, the boundary of this contour resembles the line where the magnetization is perpendicular to the background, which is also an experimentally used criterion for skyrmion size \cite{meyer2019isolated}. 

The evolution of the soliton area $A$ with increasing simulation box size $A_{\Box}$ is shown in Fig.~\ref{fig:convergency}(a). In order to overcome the limitations of our results by finite-size effects we take advantage of the observation that not only the soliton area $A$, but also the energy barrier $\Delta E$ against a collapse into the FM [Fig.~\ref{fig:convergency}(b)] and the constituents of the pre-exponential factor, namely the dynamic contribution $\Lambda^{\ddagger}$ and the entropy difference $\Delta\mathcal{S}$ as shown in Fig. \ref{fig:convergency}(c-d), can be fitted by
\begin{equation}\label{eq:fit}
    f(A_{\Box};f_{\infty},\kappa,\xi)=f_{\infty}+\frac{\kappa}{A_{\Box}^{\xi}} ~,
\end{equation}
so that the quantity for the infinite simulation box is simply the fitting variable $f_{\infty} = \lim_{A_{\Box}\to\infty}f(A_{\Box})$ if $\xi>0$. 

As shown in Fig.~\ref{fig:convergency}, proper convergence cannot be achieved for magnetic fields $B_z\leq 1.54~$T within the range of computationally reasonable system sizes. For skyrmion-like states above this field (green), where there is no $\mathrm{U}(1)$-symmetry, the situation is different, and reasonable convergence for all quantities is reached within the range of directly computed system sizes. Overall, we note that $\Lambda^{\ddagger}$ and $\Delta \mathcal{S}$ [Fig.~\ref{fig:convergency}(c–d)] converge faster with increasing simulation box size than $A$ and $\Delta E$ [Fig.~\ref{fig:convergency}(a–b)]. This indicates that finite-size effects are less pronounced in the pre-exponential factor than in the energy barrier. However, the entropy difference exhibits a very slow convergence near the critical field. Since all quantities were calculated in the limit of infinitely large simulation boxes, their convergence is guaranteed.

Interestingly, this behavior originates from the unbroken rotational symmetry, which is present below the critical magnetic field and leads to a long-range Belavin-Polyakov shape of solitons in at least one direction. In the following, using the prcedure from Eq.~(\ref{eq:fit}), all GNEB and HTST calculations are 
extrapolated to infinitely large honeycomb lattices in order to achieve fully converged physical quantities.
To the best of our knowledge, the stability and lifetime of bimerons and antibimerons have not been reported in the literature to date for any material. In the following two subsections, we address this gap.

\begin{figure}[tp]
	\centering
	\includegraphics[width=1.0\linewidth]{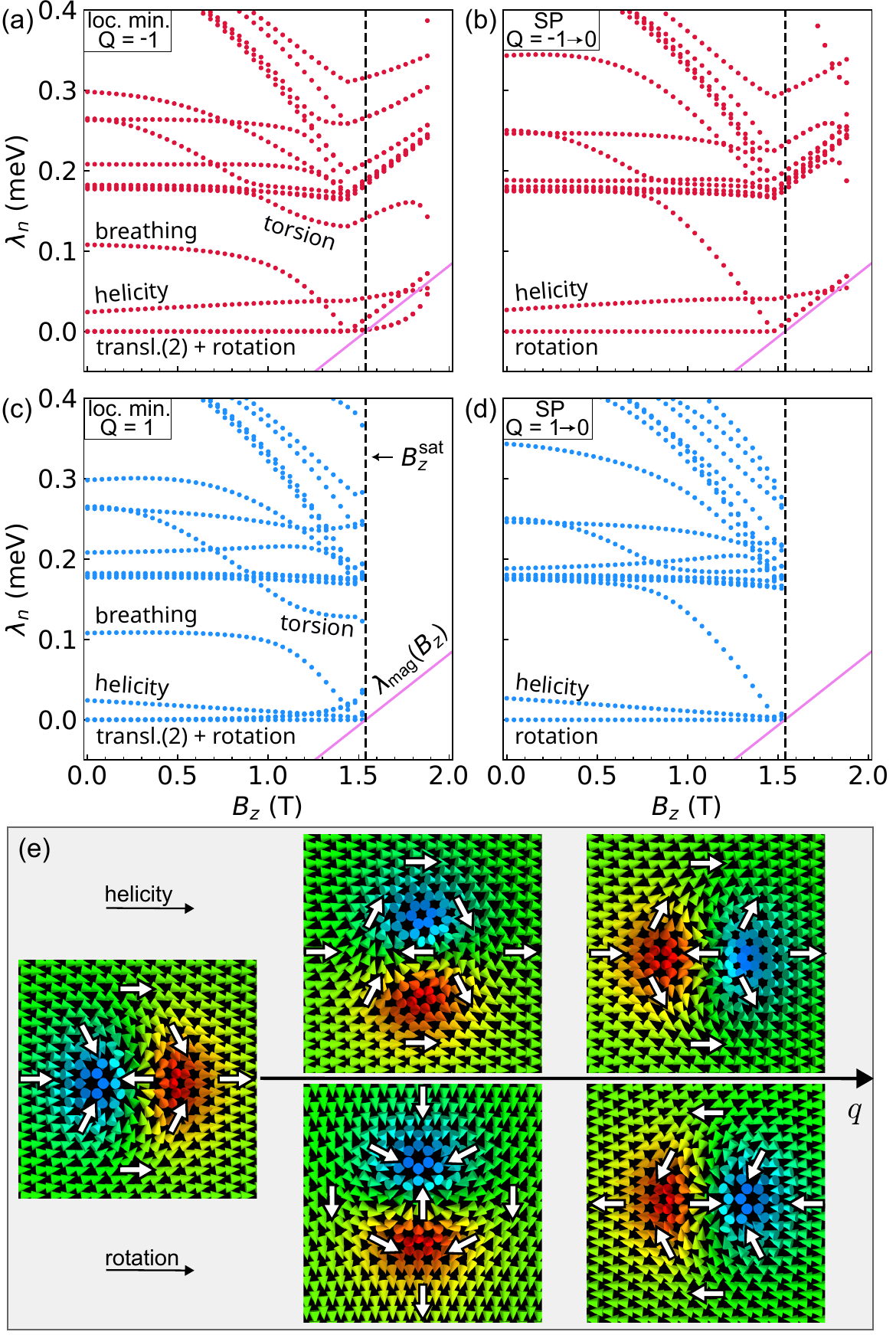}
	\caption{\label{fig:eigenvalue} Hessian eigenvalue spectra. (a) Eigenvalues $\lambda_n$ of Hessian $\mathrm{H}$ for soliton states with $Q=-1$, including bimerons and skyrmions, over magnetic fields $B_z$. (b) Eigenvalues of Hessian for the saddle point state for the collapse of $Q=-1$ state into the FM state. (c) Eigenvalues of Hessian for solitons with $Q=1$, including antibimerons and antiskyrmions, and (d) the respective saddle point for the collapse into the FM state. All eigenvalues are shown for a simulation box with $40 \times 40$ unit cells (2-atom/unit cell). The spectrum is subject to the simulation box size below $B_z^{\text{sat}}$ (vertical dashed line). In all panels, the low-energy degrees of freedom are labeled, including the zero-energy excitations corresponding to translations and rotations of the helicity.
   The magenta line shows the magnon gap [cf.~Eq.~(\ref{eq:magnon_gap})]. (e) Action of the helicity and the rotation mode on the bimeron at $B_z=0~$T along deflection magnitude $q$ [cf. Eq.~(\ref{eq:harmonic_approx})]. The central aspect of the comparison is that the helicity keeps the orientation of the background, but changes the helicity of merons, which comes along with a cost in DMI. The rotation changes the orientation of the background, but keeps the helicity of merons, which makes it a symmetry of the system.
}
\end{figure}

\subsection*{H. Entropy under unbroken rotational symmetry}

Similar to the long-range soliton profiles reported in Subsection~E, the unbroken rotational symmetry within the easy-plane allows for gapless long-range excitations of magnetic textures. Away from thermal vortex formations that reduce the spin-correlation length in a realistic 2D material, the extent of these excitations is usually bound by the size of the simulation box \cite{bramwell1994magnetization}. Regarding the magnon gap of our Hamiltonian
\begin{equation}\label{eq:magnon_gap}
    \lambda_{\text{mag}} = \mu B_z + 2K,
\end{equation}
it is clear that this is true for all fields $B_z\leq B_z^{\text{sat}}= -2K/\mu$ and thus applies to backgrounds that feature unbroken $\mathrm{U}(1)$-symmetry, as indicated by circles and cones in Fig.~\ref{fig:bimeron_skyrmion_transformation}. Due to the strong connection between symmetry and entropy, this highly affects the lifetime of solitons in this regime.

In the framework of HTST, the entropy (see Section \ref{theory}) directly depends on the local curvatures $\lambda_n$ of the energy landscape around the IN and SP, as well as the configuration space volume of symmetry equivalent states $L$, also referred to as zero-mode partition functions
\begin{equation}\label{eq:entropy}
     \frac{\Delta \mathcal{S}}{k_{\text{B}}} = \ln\left[\left(\frac{2\pi}{\beta}\right)^{\Delta k/2}\frac{L_{\text{sp}}\prod_{n=1}^{k_{\text{in}}}\sqrt{\lambda_n^{\text{in}}}}{L_{\text{in}} \prod_{n=1}^{k_{\text{sp}}}\sqrt{\lambda_n^{\text{sp}}}}\right] + \frac{\Delta k}{2} ~,
\end{equation}
with $\beta^{-1} = k_{\text{B}}T$. Here, the products run over all $k$ harmonic degrees of freedom of a given state and $\Delta k = k_{\text{sp}}-k_{\text{in}}$ marks their difference, so that the factor $(2\pi/\beta)^{\Delta k/2}$ ensures that the quantity in Eq.~(\ref{eq:entropy}) is dimensionless.

The curvatures $\lambda_n$, obtained as the eigenvalues of the spherical constraint Hessians \cite{bessarab2012harmonic} are shown in Fig.~\ref{fig:eigenvalue}. The absence of the magnon gap below $B_z^{\text{sat}}$ [cf. Eq.~(\ref{eq:magnon_gap})] means that there is no separation of long-range, magnon-like excitations and localized deflections of the soliton. Since long-range excitations are naturally affected by system size, the eigenvalue spectrum is heavily pronounced by the dimension of the simulation box [cf. APPENDIX~A.5].

However, it has to be noted that the absence of the magnon gap raises the necessity of rescaling the interactions for magnon excitations, making them explicitly size and temperature dependent. This is stated by the Berezinskii-Kosterlitz-Thouless (BKT) theory \cite{kosterlitz1973ordering, bramwell1994magnetization}, which has been found to be applicable to easy-plane Heisenberg models \cite{liao2024tunable}. Such a rescaling procedure within the HTST framework, to the author's knowledge, has not been developed yet and therefore is not included in this work. Thus, the entropy contribution to the lifetime we present in this work, which is dominantly determined by the spectrum of magnon excitations, is only meaningful below the critical temperature of the BKT transition, which effectively constitutes an additional limitation on top of the usual rare-event condition of $5 k_{\text{B}}T\lesssim \Delta E$.

In Fig.~\ref{fig:eigenvalue}, we show low-energy excitations with eigenvalues $0 \leq \lambda \leq 0.4$ meV for a simulation box of $40 \times 40$ unit cells, with 2 atoms per unit cell. Below $B_z^{\text{sat}}$, we identify three symmetries for bimerons [Fig.~\ref{fig:eigenvalue}(a)] and antibimeron [Fig.~\ref{fig:eigenvalue}(c)] in agreement with Ref.~\cite{deng2025}, which manifest as zero-energy excitations: the rotation of spin textures with respect to the lattice and the translation into two orthogonal directions. In this work, the numerical threshold for their identification is set to $|\lambda_n| \leq 1\cdot10^{-6}~$eV. The respective zero-mode partition functions are computed as described in Ref.~\cite{Haldar2018, goerzen2023lifetime}. Beyond the rotation of localized merons, this mode involves a rotation of all spins within the uniform background in the $xy$-plane, the respective partition function diverges with system size $A_{\Box}$, becoming more and more independent of the localized soliton texture and thus approaches similar values for initial state and saddle point. Although this degree of freedom marks a difference to the eigenvalue spectrum of skyrmions or antiskyrmions \cite{Malottki2019, goerzen2023lifetime, schrautzer2022effects}, the contribution of this mode to the entropy difference vanishes for $A_{\Box}\to\infty$. Note that the rotation has to be distinguished from the helicity mode \cite{lin2016ginzburg}, as depicted in Fig.~\ref{fig:eigenvalue}(e). While both modes describe a rotation of the meron pairs around their common geometric center, the helicity mode describes a continuous change to the helicity $\gamma$ in the azimuthal soliton profile function $\Phi(\phi)$ [cf. Eq.~(\ref{eq:skyrmion_ansatz})]. While this deflection leaves the orientation of the magnetic background constant, it continuously rotates the helicity of the two merons, which consequently is accompanied by an increase in DMI located at the meron with vortex number $\nu=1$, as stated by Eq.~(\ref{eq:DMI_response1}). Thus, the helicity mode has no symmetry of the system and appears as a harmonic mode in all four spectra depicted in Fig.~\ref{fig:eigenvalue}.

An important difference to the eigenvalue spectrum of skyrmions or antiskyrmions \cite{Malottki2019, goerzen2023lifetime, schrautzer2022effects} is the existence of a rotation, describing a rotation of the soliton by leaving the direction of the background unchanged, next to the usual rotation of helicity \cite{lin2016ginzburg}, which rotates all spins but leaves the meron centers unchanged. 

\subsection*{I. Soliton lifetimes}

\begin{figure}[tp]
	\centering
	\includegraphics[width=1.0\linewidth]{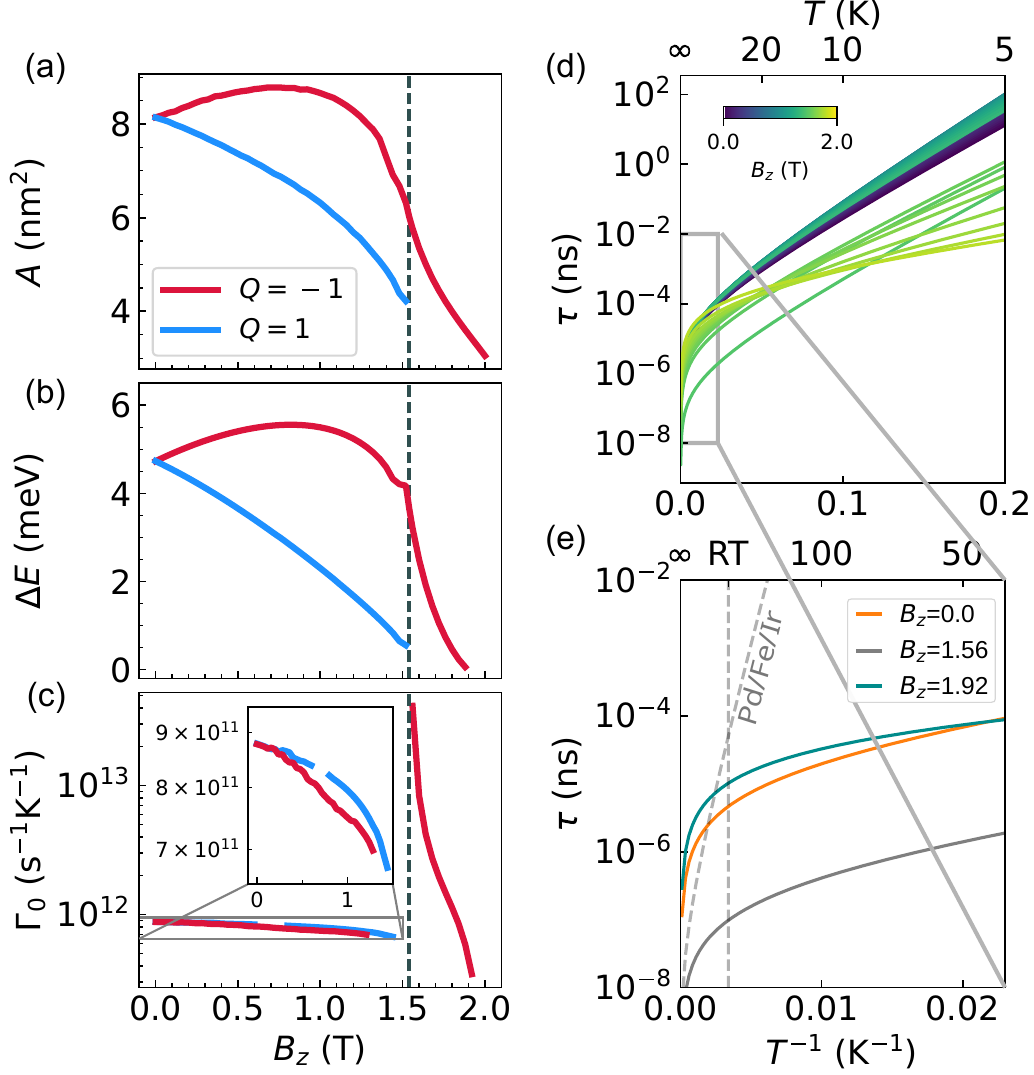}
	\caption{\label{fig:lifetime} Soliton lifetime. 
    (a) Area $A$ of solitons in background canted by magnetic field with strength $B_z$. The vertical dashed line gives the position of the saturation field $B_z^{\text{sat}} = 1.54~$T. The soliton area is extrapolated to infinitely large simulation box sizes.
    It is distinguished between solitons with $Q=-1$ and $Q=1$ solitons. Similarly, (b) shows the energy barrier $\Delta E$ and (c) the pre-exponential factor $\Gamma_{0}$ for a collapse into the field-polarized state. For visualization, the linear temperature dependence, arising from $\Delta k=1$, is removed from the pre-exponential factor, leaving it in units of s$^{-1}$K$^{-1}$.
    (d) Lifetime of solitons [cf. Eq.~(\ref{eq:log_arrhenius_law})] at different $B_z$, indicated by the color bar in the top left corner, are displayed logarithmically over inverse temperature. The temperatures themselves are given on the upper abscissa. (e) Zoom on lifetimes in the high-temperature regime. The comparison to the lifetime of skyrmions in the ultrathin film fcc-Pd/Fe/Ir(111) at $B_z=3.9~$T (dashed line) \cite{Malottki2019} shows that, despite their low energy barrier, solitons at different $B_z$ can compete with previously experimentally feasible solitons due to entropic stabilization. 
}
\end{figure}

For a quantitative analysis of soliton properties under a magnetic field-induced canting of the FM background, we start with their field-dependent size.
The evolution of the soliton size is displayed in Fig.~\ref{fig:lifetime}(b). Here, it can be observed that bimerons and antibimerons are degenerate in size at zero field, but this degeneracy is lifted for $B_z>0~$T. In particular, we observe that solitons with $Q=-1$ grow in size, while those with $Q=1$ shrink. 
These numerical results are in perfect agreement with the theoretical prediction in Fig.~\ref{fig:bimeron_skyrmion_transformation}, where the splitting is linked to differences in response to DMI.

Similarly, as for $B_z=0~$T (see Fig.~\ref{fig:bimeron_mep}), we perform MEP calculations for solitons with $Q=\pm1$ 
from $B_z=0~$T to $B_z=2~$T. From the obtained SP, we compute the activation energy $\Delta E$ as well as the pre-exponential factor $\Gamma_{0}$ 
[cf.~Eq.~(\ref{eq:Arrhenius})], both for infinitely large lattices.
The energy barrier $\Delta E$ follows the trend of soliton size as previously found for skyrmions \cite{varentcova2020toward}.
A similar, yet inverted progression can be observed for the pre-exponential factors in the zoomed inset in Fig.~\ref{fig:lifetime}(c). It is computed from the limit of the dynamical factor and the entropy difference [cf. Fig.~\ref{fig:convergency}(c-d)] following Eq.~(\ref{eq:arrhenius_law}). Considering $\Delta k=1$, following from the number of harmonic degrees of freedom discussed in the previous section, it follows that $\Gamma_0\propto T$, so that it is displayed temperature independent in units of s$^{-1}$K$^{-1}$ in Fig.~\ref{fig:lifetime}(c). Here, the pre-exponential factors for solitons with $Q=1$ are found to be slightly higher than for $Q=-1$ for $B_z>0~$T. However, the magnitude of this difference vanishes in the face of the variations of pre-exponential factors for $B_z>B_z^{\text{sat}}$. Phenomenologically, this huge difference between the two regimes coincides with the emergence of a magnon gap and thus seems to stem from contributions from long-range magnon-like excitations.

It has also to be noted that the breathing mode in Fig.~\ref{fig:eigenvalue}, which describes a change in the size of the soliton, becomes a zero mode near $B_z^{\text{sat}}$. This is in good agreement with the undefined size of Belavin-Polyakov solitons \cite{polyakov22metastable, ivanov2007quantum} obtained in situations of critical coupling \cite{zarzuela2020stability}. Note that despite the deviation between the numerically obtained root of the breathing mode, visible in Fig.~\ref{fig:eigenvalue}(a)-(d), and $B_z^{\text{sat}}$, we find that the difference decreases with system size [cf. Fig.~\ref{fig:spectrum_size_dependence}(a),(b)], suggesting an agreement between both for infinitely large lattices. However, practically, it hinders us from computing reasonable entropy differences, and therefore pre-exponential factors, in a certain region before reaching the critical point.

In Fig.~\ref{fig:lifetime}(d), the soliton lifetime with $Q=-1$ for different magnetic fields is displayed over inverse temperature. Fig.~\ref{fig:lifetime}(e) shows a zoom on lifetimes at particular magnetic fields, which compares their behavior at asymptotically high temperatures. Although the lifetimes are on an overall low level, the comparison to
skyrmions in Pd/Fe/Ir(111) (dashed) indicates comparable lifetimes at sufficiently high temperatures. The origin of this effect is that the soliton lifetimes 
have a higher offset at $T^{-1}\to0$ compared to skyrmions in Pd/Fe/Ir(111).
Considering the logarithmic Arrhenius law
\begin{equation}\label{eq:log_arrhenius_law}
    \ln \tau = \beta\Delta E - \ln\Gamma_0~.
\end{equation}
this offset is determined by $\Gamma_{0}$ and therefore related to entropic stabilization of solitons (see Section \ref{theory}). In the same way, the slope of lifetimes is determined by the energy barrier $\Delta E$. On the one hand, this explains the superior lifetime of skyrmions in Pd/Fe/Ir(111) for low temperatures since solitons in CGT have comparably small energy barriers of $\Delta E < 6~$meV. On the other hand, since energy barriers dominantly define 
the lifetime's response to variations in temperature, it reveals that the lifetimes of solitons in CGT are significantly less affected by 
thermal fluctuations. Although the rare-event-condition $5 k_{\text{B}}T\lesssim \Delta E$ states the validity of HTST up to $T\lesssim 11~$K, this insight is explicitly important in combination with the pre-exponential factors themselves being only marginally influenced by magnetic fields $B_z\leq 1.54~$T. This makes entropy-stabilized bimerons and antibimerons in 2D materials a promising platform for spintronic applications, which are supposed to be robust against fluctuations of environmental influences.

\section{CONCLUSIONS AND OUTLOOK}
In this combined study between DFT and HTST, we investigated the lifetimes of meron-bound states with topological charges $Q=\pm1$ in the CGT layer of the FGT/CGT interface and observed the evolution of their stability under an applied magnetic field perpendicular to the easy-plane of CGT. We distinguish two regimes separated by the critical magnetic field $B_z^{\text{sat}}=1.54~$T. Below this field, the rotation within the easy-plane remains an unbroken symmetry, which affects the stability and shape of solitons, which range from degenerate bimerons and antibimerons to non-degenerate skyrmions and antiskyrmions. While the unbroken symmetry leads to Belavin-Polyakov-like long-range profiles of solitons, it also hinders the formation of a magnon gap, which in turn means that the entropy is dominated by contributions from non-local excitations.
We further conclude that the entropic stabilization of magnetic solitons in CGT exceeds that in the heavily investigated ultrathin film Pd/Fe/Ir(111), visible from Fig.~\ref{fig:lifetime}(e).
Since entropy is the leading state function for lifetimes beyond ultra-cold temperatures, we emphasize the viability of easy-plane solitons in 2D vdW magnets for future computing schemes that utilize more than one kind of soliton. In this regard, it has to be mentioned that the validity of the lifetimes presented in this work is restricted by two limitations: One is the rare-event condition for transitions, while the other is the BKT transition temperature, near which the rescaling of magnetic interaction due to increased meron-antimeron pair nucleations might alter the entropic contributions not negligibly. An extension of the HTST framework including BKT physics is still subject to future developments.

From a materials perspective, the proposed 2D vdW heterostructures can serve as a promising platform to realize a wider variety of magnetic solitons with distinctive symmetry and topology \cite{zhu2026strongly}, whose stability can be easily tuned by external stimuli such as gating, strain, or twistronics \cite{burch2018magnetism, Dongzhe_prb2023,schwab2025skyrmion}.

For future device applications, our proposal is beneficial for neuromorphic and stochastic spintronic devices \cite{pinna2018, song2020}, which require a high degree of non-linear soliton-soliton interactions, or in magnetic soliton-based quantum computing \cite{skyquibit_2021}, 
Therefore, the stabilization mechanism we propose is general for easy-plane solitons with arbitrary $Q$, not limited to $Q = \pm1$, but also applicable to higher-order ($|Q|>1$) solitons.
Thus, it can be expected that the stability of high-$Q$ states in easy-plane magnets fundamentally exceeds that in easy-axis magnets because of their different response to broken inversion symmetry.
This is in strong contrast to DMI-stabilized easy-axis solitons, where skyrmions are stabilized by interfacial \cite{goerzen2023lifetime, kuchkin2020turning} and antiskyrmions by anisotropic DMI \cite{hoffmann2017antiskyrmions}. However, since the additional entropic stability for solitons in easy-plane magnets stems from the unbroken $\mathrm{U}(1)$-symmetry of the lattice, the robustness of this result under the addition of interactions that lift this symmetry, such as 3-site exchange interactions or 3-fold in-plane anisotropy on the honeycomb, is still an aspect of investigation \cite{schrautzer2026impact}. Finally, we emphasize that the proposed mechanism in this work is general and should apply to various systems such as ultrathin films, magnetic multilayers, and vdW interfaces. 
By optimizing relevant magnetic interactions, particularly by combining the entropy-dominated lifetime with large energy barriers, taylored by system geometry under unbroken $\mathrm{U}(1)$-symmetry, we expect advances in soliton technology toward room-temperature operation.

\section*{APPENDIX A: THEORETICAL DETAILS}
\label{methods}

\subsection*{1. Computational details of the DFT calculations}

Our first-principles calculations were performed using the \textsc{Fleur} code \cite{fleurv26} based on the full-potential linearized augmented plane wave method (FLAPW). We used local density approximation in the 
parametrization for exchange-correlation functionals
of Vosko, Wilk, and Nusair \cite{vosko1980accurate}. The CGT/FGT heterostructure consists of 1 monolayer of CGT coated by 1 monolayer of FGT. The in-plane lattice constants of CGT and FGT are 6.93 \AA~and 4.00 \AA, respectively. A  $(1 \times 1)$  CGT unit cell was matched to a 
$(\sqrt{3} \times \sqrt{3})$ FGT cell by fixing the FGT lattice constant. The CGT lattice is slightly stretched by about 0.15\% by forming the CGT/FGT interface. To describe the vdW distance between CGT and FGT accurately, we also included in our calculations vdW interactions using semi-empirical dispersion corrections (DFT-D3) as formulated by Grimme \cite{grimme2010consistent}. The interlayer vdW distance was found to be 3.78 \AA, indicating a rather weak interlayer interaction. To treat the on-site correlation properly, we applied LDA+$U$ with $U_{\text{eff}} = 0.5$ eV for CGT, close to the one used in Ref. \cite{gong2017discovery}, reproducing the correct experimental magnetic ground state. In contrast, we have not considered the Hubbard $U$ correction for FGT since LDA yields a better magnetic moment for the Fe atom closer to the experimental one, as previously pointed out in Refs. \cite{Zhuang2016,deng2018gate}.

To obtain the pair-wise Heisenberg exchange and DMI constants required for the atomistic spin model [cf. Eq.~(\ref{eq:energy_model})], we have performed spin spiral calculations via DFT \cite{Kurz2004,Heide2009,dupe2014tailoring}.
A spin spiral is characterized by a wave
vector $\mathbf{q}$, and the spin at lattice site $\textbf{r}_i$ is given by $\mathbf{m}_i=(\cos(\mathbf{q}\cdot \mathbf{r}_i),\sin(\mathbf{q}\cdot \mathbf{r}_i),0)$.
The energy dispersion 
neglecting SOC, $E_{\text{ss}}(\V{q})$, is calculated via the {\tt FLEUR} code
in the scalar-relativistic approximation
based on the generalized Bloch theorem \cite{Kurz2004}.
The Heisenberg exchange constants are obtained by fitting the
self-consistently converged homogeneous flat spin spiral dispersions without SOC, $E_{\text{ss}}(\V{q})$
\begin{equation}\label{dft1}
E_{\text{ss}}(\mathbf{q})
= - \sum_{\sigma} J_{\sigma} \sum_{j} \cos\!\left(\mathbf{q} \cdot \mathbf{r}_{j}\right),
\end{equation}
Here, $\sigma$ labels the coordination shell, and the index $j$ runs over the
symmetry-equivalent neighbors within shell $\sigma$. The vectors
$\mathbf{r}_{j}$ connect a reference magnetic site to its neighbors, such that
all $|\mathbf{r}_{j}|$ are equal within a given shell. The parameters
$J_{\sigma}$ denote the corresponding shell–resolved exchange constants.

The DMI contribution to the dispersion, $E_{\rm DMI}(\mathbf{q})$,
is computed within first-order perturbation theory \cite{Heide2009} as follows
\begin{equation}\label{dft2}
\begin{split}
E_{\mathrm{DMI}}(\mathbf{q}) &=
 \sum_{\mathbf{k} \nu} n_{\mathbf{k} \nu}(\mathbf{q})\,
 \delta \epsilon_{\mathbf{k} \nu}(\mathbf{q}), \\
\text{where } \delta \epsilon_{\mathbf{k} \nu}(\mathbf{q}) &=
 \left\langle\psi_{\mathbf{k} \nu}(\mathbf{q})\right|
 H_{\mathrm{SOC}}
 \left|\psi_{\mathbf{k} \nu}(\mathbf{q})\right\rangle .
\end{split}
\end{equation}
where $\psi_{\mathbf{k} \nu}$ denotes scalar-relativistic spin-spiral eigenstates (without SOC), and $n_{\mathbf{k} \nu}$ is the occupation number of the eigenstate $\psi_{\mathbf{k} \nu}$. The total energies in 
Eq.~(\ref{dft1}) and 
Eq.~(\ref{dft2}) are then mapped to the last term of 
Eq.~(\ref{eq:energy_model}) in order to obtain exchange and DMI constants in arbitrary NN.

Note that we have used an extended BZ beyond the first BZ in order to describe accurately the magnetic interactions in the complex geometrical structure of the interface (see the following 
Subsection 2 for more details). A cutoff parameter for FLAPW basis functions of $k_{\text{max}}$ = 4.1 a.u.$^{-1}$ was used in our calculations. We have converged the total energy of flat spin spiral states using a $21 \times 21$ $\V{k}$-points mesh in order to obtain an accuracy of 0.001 meV.

Finally, the MAE (single-ion anisotropy constant $K$) was calculated using the band-energy difference (i.e., the so-called magnetic force theorem)
\begin{equation}\label{dft3}
K = \sum_{\alpha \in \mathrm{occ.}} 
\left( \epsilon_{\alpha}^{\parallel} - \epsilon_{\alpha}^{\perp} \right),
\end{equation}
where the summation runs over the occupied electronic states. Here, $\epsilon_\alpha$ are the eigenvalues obtained from a single diagonalization of the Hamiltonian including SOC, starting from self-consistently converged charge densities with rotations to the appropriate spin-orientation axis. We used a denser $\V{k}$-point mesh of $87 \times 87$ in the full 2D BZ for these calculations.

\subsection*{2. Spin model parameters: Mapping from the hexagonal to the honeycomb lattice}

For CGT, the magnetic Cr atoms, which are solely relevant for atomistic spin simulations, form a honeycomb lattice. To simplify DFT calculations, we treated the two Cr atoms in CGT collectively, meaning that all spin-spiral calculations were performed by rotating both Cr atoms simultaneously. By fitting the spin-spiral dispersion curves in Fig. \ref{spin-spiral}(c) and (d) of the main text, we directly extract the magnetic interaction parameters for the hexagonal lattice.

\begin{figure}[b]
    \centering
    \includegraphics[width=1.0\linewidth]{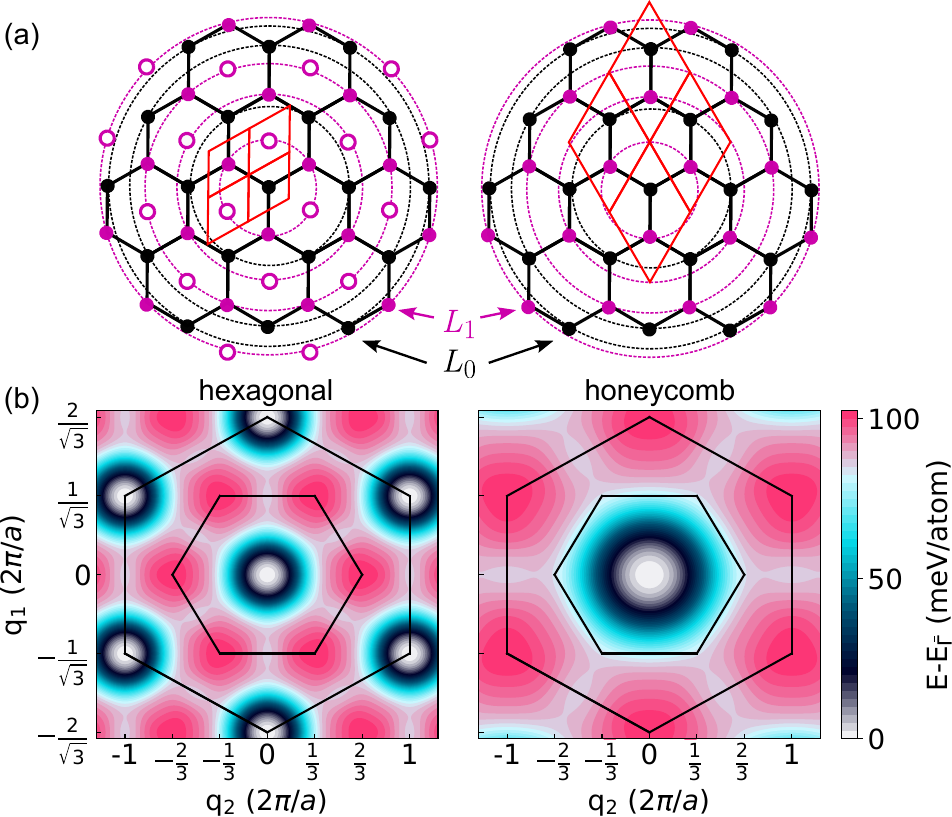}
    \caption{Energy on reciprocal honeycomb and hexagonal lattices. (a) The hexagonal lattice can be decomposed into three hexagonal lattices (black, violet, hollow dots), each with lattice constant $a=1$. Removing one of them, here the hollowed dots, leaves a honeycomb lattice with a larger unit cell (red lines) of $\tilde{a}=\sqrt{3}$ and rotated by $\pi/6$. (b) Exchange energy with respect to the $\Gamma$-point for the reciprocal hexagonal (left) and honeycomb (right) lattice models of CGT.
It is illustrated that the honeycomb BZ, compared to the hexagonal BZ, is rotated by $\pi/6$ and scaled by $\sqrt{3}$. This is caused by the necessity of displaying the reciprocal dispersion in $\mathbf{q}$-spaces of the same scale, which requires shrinking and rotating the real-space honeycomb unit cell. Due to the inverse relation between length scales in real and reciprocal spaces, this causes a rotation and expansion of the BZ.} 
    \label{fig:honeycomb}
\end{figure}

The honeycomb lattice can be formed from two hexagonal lattices, as shown in 
Fig.~\ref{fig:honeycomb}(a). This means that a set of magnetic interaction constants can be transformed so that the energy over the respective reciprocal lattices of the hexagonal and honeycomb lattices remains similar. A mapping between these two lattices is needed. The concept behind this mapping is illustrated in 
Fig.~\ref{fig:honeycomb}(b), where the energy dispersion within the first BZ of the honeycomb lattice (left panel, outer hexagon) coincides with the dispersion within the first BZ of the hexagonal lattice (right panel, inner hexagon). 

To understand the transformation of the interaction constants that lead to this convergence between lattice geometries, consider that the rather complicated honeycomb lattice can be modeled from a hexagonal lattice by adding additional lattice sites in the center of each hexagon, as illustrated by the grey points in Fig.~\ref{fig:honeycomb}(a). Further, each of these virtually added lattice sites at position $\V{r}'$ has a non-virtual counterpart at $\V{r}=-\V{r}'$. This has an important implication for spin spiral states, which are characterized by the vectors $\V{R}, \V{I}, \V{q}$ $\in$ $\mathbb{R}^3$. In this construction, $\mathbf{q}$ describes the direction in which the spin spiral propagates, while the length $|\mathbf{q}|$ is inversely proportional to its wavelength. On the other hand, the vectors $\V{R}, \V{I}$ span the plane of rotation, so that the orientation of a spin $\mathbf{m}_i$ at position $\mathbf{r}_i$ reads
\begin{equation}
    \mathbf{m}_i = \V{R}\cos(\mathbf{q}\cdot\mathbf{r}_i) - \V{I}\sin(\mathbf{q}\cdot\mathbf{r}_i)~.
\end{equation}
The energy of such a spin spiral $\mathbf{m}=\{\mathbf{m}_1,...\mathbf{m}_N\}$ reads
\begin{align}
\frac{E(\mathbf{m})}{N} 
&= -\sum_\sigma J_\sigma \sum_{\mathbf{r} \in\sigma} 
       \cos(\mathbf{q} \cdot \mathbf{r})  \notag \\
&\quad - 2 \mathbf{I} \times \mathbf{R} 
       \sum_\sigma D_\sigma \sum_{\mathbf{r} \in \sigma} 
       (\mathbf{r} \times \hat{\mathbf{z}}) \sin(\mathbf{q} \cdot \mathbf{r}) 
       - \frac{K}{2}.
\end{align}
from which it can be seen that their exchange interactions, DMI, and MAE are invariant under the inversion of spatial coordinates $\V{r} \rightarrow -\V{r}'$. This implies that within a spin spiral state, all virtually added lattice points have the same energy as their nonvirtual counterparts.

Since the virtual lattice sites only appear on certain shells, specifically the magenta 
shells in Fig. \ref{fig:honeycomb}(a), these shells will contribute twice the energy on the hexagonal lattice
compared with their contribution to the honeycomb lattice. For a distinct notation, we can use the fact that the twice-interacting shells $\sigma$ with lattice sites $\V{r} \in \sigma$ are not subsets of the sublattice $L_0$ that contains the center at $\V{r} = 0$.  In 
Fig.~\ref{fig:honeycomb}(a), this is reflected by the fact that hollow lattice sites never lie on black circles, at least within the range of investigated shells. With that, the double contribution can be treated effectively by scaling the interaction constants according to
\begin{equation}
    \arraycolsep=1.6pt\def\arraystretch{1.4}
    C_{\sigma}^{\text{hon}} = \left\{\begin{array}{cl}
        2C_{\sigma}^{\text{hex}} &~\text{if } \sigma \in L_1\\
        C_{\sigma}^{\text{hex}} &~ \text{if } \sigma\in L_0
    \end{array}\right.~,
\end{equation}
where $C_{\sigma}^{\text{hex}} \in {J_{\sigma}, D_{\sigma}}$ refers to the constants in the hexagonal lattice. 

For example, the interaction between nearest neighbors on the honeycomb lattice has 3 atoms and an exchange constant of $J_1^\text{hon}$). A corresponding hexagonal lattice with 6 nearest neighbors and an exchange constant of $J_1^\text{hex}=\frac{1}{2}J_1^\text{hon}$ 
with a weighting factor of $\frac{1}{2}$ 
would produce the same energy contributions for all spin spirals.
The shell containing all next-nearest neighbors in the honeycomb lattice is already centrosymmetric. Therefore, the weighting factor becomes 1, and we find $J_2^\text{hex}=J_2^\text{hon}$. Exchange constants and the absolute value of the DMI vector always share the same weighting factor,
which is always $\frac{1}{2}$ or 1, based on whether the shell of neighbors is centrosymmetric or not. The interaction parameters for the honeycomb lattice in Table 1 in the main text could be calculated by fitting with the dense hexagonal lattice and determining the weighting factors. Note that, as we have checked, with these two sets of parameters for CGT-hexagonal and CGT-honeycomb, our atomistic spin simulations predict similarly shaped bimerons at $B_z=0$~T that only differ by a factor $\sim\sqrt{3}$ in size.

\subsection*{3. Soliton model}
Each soliton is associated with a field of spins $\mathbf{m}:~\mathbb{R}^2\to\mathbb{S}^2$ at discrete lattice sites $\mathbf{r}$, which are initialized as
\begin{equation}\label{eq:skyrmion_ansatz}
    \mathbf{m}(\mathbf{r}) = \left( \begin{array}{c}    
    \cos\Phi(\mathbf{r})\sin\Theta(\mathbf{r})\\
    \sin\Phi(\mathbf{r})\sin\Theta(\mathbf{r})\\
    \cos\Theta(\mathbf{r})
    \end{array}  \right)~,~ \Phi(\mathbf{r}) = \nu\phi(\mathbf{r})+\gamma ~.
\end{equation}
where the azimuthal profile function $\Phi$ depends on the vorticity $\nu\in\mathbb{Z}$ and helicity $\gamma\in[0,2\pi]$. The radial $\Theta$-profile is assumed to be either exponentially or algebraically decaying [cf. Eq.~(\ref{eq:theta_profile})] with radial distance $\rho$ from the soliton center. In agreement with Eq.~(15) we define $\Theta$ with respect to the polarized background $\mathbf{m}_{\text{FM}}$. Considering $\mathrm{R}\in\mathrm{SO}(3)$ with $\mathbf{m}_{\text{FM}} = \mathrm{R}\hat{\mathbf{z}}$ the argument $\phi(\mathbf{r})=\text{arctan2}(\mathbf{r}\cdot\mathbf{y}', \mathbf{r}\cdot\mathbf{x}')$ for azimuthal component $\Phi(\phi)$ has to be defined in the rotated basis $(\mathbf{x}', \mathbf{y}', \mathbf{m}_{\text{FM}})$ with $\mathbf{x}'=\mathrm{R}\mathbf{x}$, $\mathbf{y}'=\mathrm{R}\mathbf{y}$.
While the exponential profile is motivated phenomenologically \cite{bocdanov1994properties} and describes the skyrmion as a cyclic domain wall, the algebraic profile is the analytic solution to the continuum model \cite{barton2020magnetic} if the potential energy can be brought to a form, that is similar to the situation at $B_z^{\text{sat}}$ mentioned in the main text. After initialization of the solitons, the spin textures were relaxed with respect to energy by the velocity projection optimization algorithm \cite{bessarab2015method}.

\begin{figure}[b]
    \centering
    \includegraphics[scale=0.8]{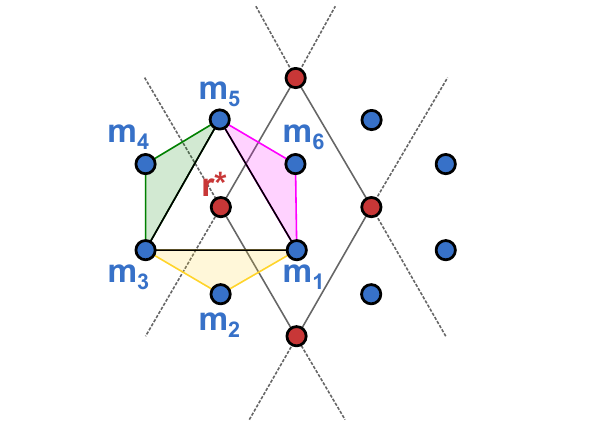}
    \caption{Triangulation of the honeycomb lattice.
    The chosen triangulation of the hexagons of the honeycomb lattice is presented for one exemplary dual lattice point $\mathbf{r}^*$ (red). The positions of the magnetic moments are represented in blue. The dual lattice corresponds to a hexagonal lattice, schematically indicated by the lines.}
    \label{fig:topodens}
\end{figure}

\begin{figure*}[t]
    \centering
    \includegraphics[width=1\linewidth]{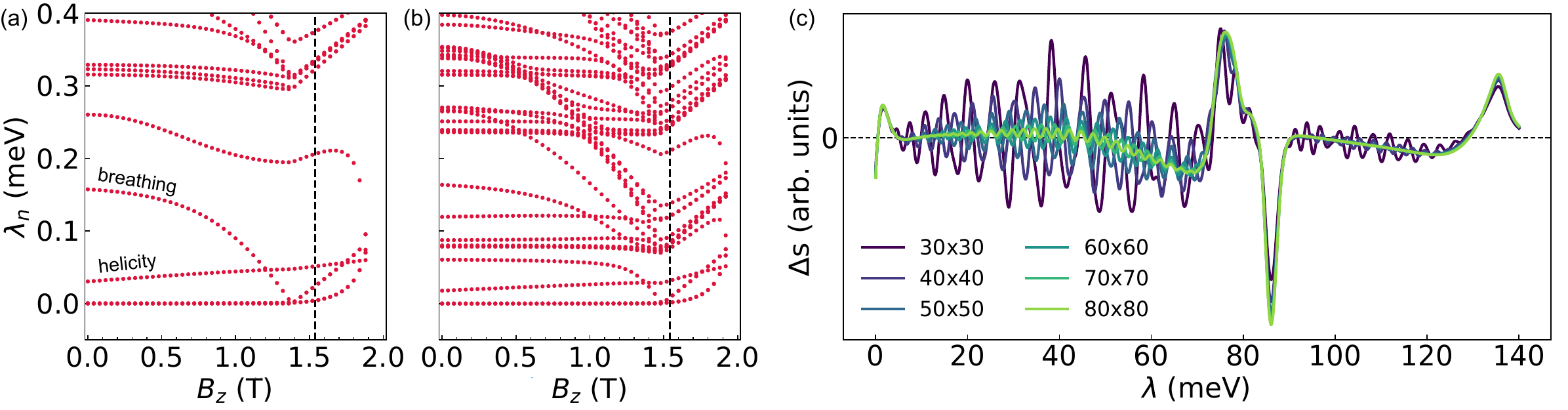}
    \caption{Dependence of Hessian eigenvalue distribution on system size. (a) Partial Hessian eigenvalue spectrum for low energy excitations of the bimeron at $B_z=0.0~$T on a lattice with $30\times30$ unit cells. The section shown is similar to those shown in Fig.~\ref{fig:eigenvalue}. (b) Partial Hessian eigenvalue spectrum of the bimeron at $B_z=0.0~$T on a lattice with $60\times60$ unit cells. The shown range of eigenvalues $0\leq \lambda \leq 0.4$ is the same as in (a). (c) Spectral entropy density difference $\Delta s(\lambda)$, as defined in Eq.~(\ref{eq:entropy_density}), evaluated for bimerons at $B_z=0.0~$T in 6 simulation boxes of different sizes. The density is computed for the full spectrum obtained by diagonalization of the Hessian, containing 3600 eigenvalues for $30\times30$ unit cells, up to 25600 eigenvalues for $80\times80$ unit cells.}
    \label{fig:spectrum_size_dependence}
\end{figure*}

\subsection*{4. Topological charge density calculation}

To characterize magnetic solitons, we calculate the topological charge $Q$ as the sum over the topological density $q$ of selected spin structures in the honeycomb lattice of CGT.
For a discrete model, the topological charge from Eq.~(\ref{eq:topological_charge}) has the form~\cite{berg1981}:
\begin{equation}\label{eq:topological_charge}
    Q=\sum\limits_{\mathbf{r^*}}q(\mathbf{r}^*)~,
\end{equation}
where the topological density $q(\mathbf{r}^*)$ is defined at the vertices $\mathbf{r}^*$ of the dual lattice.

For the honeycomb lattice of the CGT layer, six spins are adjacent to the dual lattice point $\mathbf{r}^*$. Thus, a triangulation can be used to define the spherical triangles. We use a Delaunay triangulation, yielding one large main triangle and three smaller triangles. For each of these triangles, the topological density can be computed and assigned to the dual lattice points of the triangulated lattice. The topological density assigned to the center $\mathbf{r}^*$ of the hexagons is then:
\begin{align}
\begin{split}
    q(\mathbf{r}^*)=\frac{1}{4\pi}(&\mathcal{A}(\mathbf{m}_1,\mathbf{m}_5,\mathbf{m}_3)+\mathcal{A}(\mathbf{m}_1,\mathbf{m}_3,\mathbf{m}_2)\\
    &+\mathcal{A}(\mathbf{m}_3,\mathbf{m}_5,\mathbf{m}_4)+\mathcal{A}(\mathbf{m}_5,\mathbf{m}_1,\mathbf{m}_6))~.
\end{split}
\end{align}
For the definition of the triangulation as well as for the indexation of the magnetic moments in the formulas above, as shown in Fig. \ref{fig:topodens}. 

It is worth noting that, to obtain a smooth visualization in Fig.~\ref{fig:bimeron_mep}(b) in the main text, the density was plotted at the midpoints of all four triangles forming a hexagon but normalized with respect to the area of these triangles formed by sites on the honeycomb lattice.

\begin{table*}[t]
	\centering
	\scalebox{1.0}{
		\begin{tabular}{ccccccccccccccccccc}
			\hline\hline
& $J_1$ & $J_2$ & $J_3$ & $J_4$ & $J_5$ & $J_6$ & $J_7$ & $J_8$ \\ 
\hline
~~FGT/CGT~~ & ~~72.04 ~~ & ~~$-0.01$~~ & ~~$-2.00$~~ & ~~$-0.16$~~ & ~~3.08~~ & ~~$-0.56$~~ & ~~$-0.009$~~ & ~~$-0.33$~~  \\
~~Interlayer interaction~~ & $0.25$ & -1.10 & 2.27 & 0.91 & $-3.16$ & $-1.94$ & 2.08 & 1.29  \\

			\hline
	\end{tabular}}
    \caption{Total and interlayer interaction constants. Shell-resolved Heisenberg $J_i$ for FGT/CGT (including both intra- and interlayer interactions) and interlayer contributions ($J_{\text{inter}}^{\perp, \sigma}$) between the FGT and CGT layers. Note that all parameters are treated within a collective 2D spin model (i.e., eleven magnetic atoms, including nine Fe and two Cr atoms in the supercell, are treated as a whole). All values are given in meV per unit cell.} \label{table_interlayer}
\end{table*}

\subsection*{5. Size scaling of Hessian eigenvalue spectrum and entropy}
In our work, we extrapolate the entropy to infinite lattice sites by the fitting procedure described in Sec.~III.G. In order to determine the validity of this approach, we examine the spectra of all states for different lattice sizes. Exemplarily, we show the bimeron spectrum at $B_z=0.0~$T for $30\times30$ unit cells in Fig.~\ref{fig:spectrum_size_dependence}(a) and for $60\times60$ unit cells in (b). It is observed that the density of the spectrum is increased for the larger system. Considering that the number of Hessian eigenvalues in an $N\times N$ honeycomb lattice is given by $4N^2$, the increase in the number of modes is just natural. However, beyond that, a shift of some eigenvalues can be observed as well, e.g., visible by comparing the helicity modes for both system sizes in Fig.~\ref{fig:spectrum_size_dependence}, indicating that the spectral distribution of excitations does not follow a simple rule. We attempt the quantification by rewriting the harmonic contribution to the entropy in integral form
\begin{equation}
    \frac{\Delta S'}{k_{\text{B}}}= \ln\left(\frac{\prod_{n=1}^{k_{\text{in}}}\sqrt{\lambda_{n}^{\text{in}}}}{\prod_{n=1}^{k_{\text{sp}}}\sqrt{\lambda_{n}^{\text{sp}}}}\right) = \int_{\epsilon}^{\infty}\Delta s(\lambda)~\mathrm{d}\lambda~,
\end{equation}
starting from the zero-mode limit $\epsilon$. Here we introduce the integrand as the spectral entropy density function $\Delta s$
\begin{equation}\label{eq:entropy_density}
    \Delta s(\lambda) = \frac{\ln\lambda}{2}\left[\rho_{\text{in}}(\lambda) - \rho_{\text{sp}}(\lambda)\right]~,~ \rho_{\text{x}}(\lambda) = \sum_{n=1}^{k_{\text{x}}}\delta(\lambda-\lambda_n^{\text{x}})~.
\end{equation}
Approximating the $\delta$-distribution by a Cauchy distribution function with width $\sigma=2~$meV we show the entropy density function for bimerons at $B_z=0.0~$T on lattices of different size in Fig.~\ref{fig:spectrum_size_dependence}(c). While we find significant oscillations in $\Delta s$ for small system sizes (violet), these oscillations decline for larger lattices (green), approaching a smooth function.

This means that while spectral changes for increased lattice sizes are hard to quantify on a microscopic scale -- an analysis of single eigenvalue modifications -- the definition of entropy and entropy density allows for a quantification on a macroscopic scale. Since we find by numerical means that the entropy density converges to a fixed function, we argue that the extrapolation of the entropy to infinite lattice sizes does not neglect divergent effects in the excitation spectrum that appear under rescaling of the lattice.

\section*{APPENDIX B: INTERLAYER INTERACTIONS}

\begin{figure}[t]
    \centering
    \includegraphics[width=1.0\linewidth]{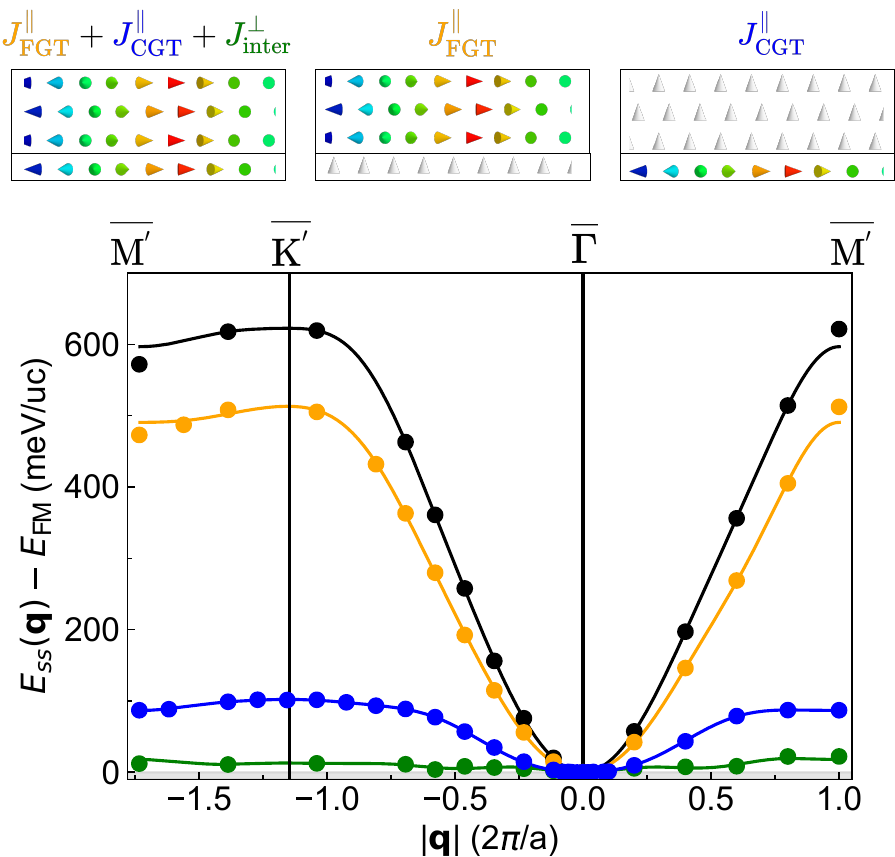}
    \caption{Total, intralayer, and interlayer interactions at FGT/CGT. The total spin-spiral curve (black) corresponds to rotating the spins of both the FGT and CGT layers ($J_{\text{FGT}}^{\parallel}+J_{\text{CGT}}^{\parallel}+J_{\text{inter}}^{\perp}$, see Eq. (\ref{interlayer})). Here, $J_{\text{FGT}}^{\parallel}$ and $J_{\text{CGT}}^{\parallel}$ denote the intralayer contributions for the FGT and CGT layers, respectively. The interlayer contribution ($J_{\text{inter}}^{\perp}$, green), corresponds to the Heisenberg exchange between the FGT and CGT layers. It is clear that the interlayer exchange is rather small, as also visible in Table \ref{table_interlayer}. On top, we show sketches of flat cycloidal spin spirals propagating in a magnetic bilayer along the direction of the wave vector $\mathbf{q}$.} 
    \label{fig:interlayer}
\end{figure}

Fig.~\ref{fig:interlayer} presents the spin-spiral dispersion of the FGT/CGT bilayer, decomposed into intralayer and interlayer exchange contributions. The total energy of the bilayer is modeled within a shell-resolved Heisenberg framework as
\begin{equation}\label{interlayer}
\begin{aligned}
E_{\text{FGT/CGT}}(\mathbf{q}) = &- \sum_{\sigma} J^{\parallel, \sigma}_{\text{FGT}} \sum_{j} \cos(\mathbf{q} \cdot \mathbf{r}_{j}) \\
&- \sum_{\sigma} J^{\parallel, \sigma}_{\text{CGT}} \sum_{j} \cos(\mathbf{q} \cdot \mathbf{r}_{j}) \\
&- \sum_{\sigma} J^{\perp, \sigma}_{\text{inter}} \sum_{j} \cos(\mathbf{q} \cdot \mathbf{r}_{j}),
\end{aligned}
\end{equation}
where the first sum $\sigma$ is the shell index, and in the second sum $j$ denotes the different sites within shell $\sigma$. Here,  magnetic moments $\mathbf{m}_j(\mathbf{q})$ are sorted in shells under the condition that the distance $|\mathbf{r}_j - \mathbf{r}_i|$ is the same for every $j$ in the same shell. The first two terms ($J_{\text{FGT}}^{\parallel, \sigma}$, $J_{\text{CGT}}^{\parallel, \sigma}$) describe the shell-resolved intralayer magnetic interactions in the FGT and CGT layers, while the last term ($J_{\text{inter}}^{\perp, \sigma}$) corresponds to the interlayer interactions between FGT and CGT.

We plot the total spin spiral dispersion (black curve in Fig. \ref{fig:interlayer}) for FGT/CGT, meaning that the spin spiral propagates in both the FGT and CGT layers. By mapping these total energies onto Eq.~(\ref{interlayer}), one obtains exchange constants for FGT/CGT as summarized in Table \ref{table_interlayer}. The decomposition into individual contributions (orange: FGT, blue: CGT, green: interlayer) shows that the dispersion is dominated by intralayer interactions, with the FGT layer providing the largest contribution.
The top panels in Fig. \ref{fig:interlayer} illustrate representative flat cycloidal spin spirals propagating along $\mathbf{q}$, emphasizing that the extracted dispersion corresponds to homogeneous rotations within each layer.

In contrast, the interlayer exchange term remains nearly flat and two orders of magnitude smaller, indicating a weak and short-ranged coupling between the layers. This behavior is consistent with the vdW character of the interface. The negligible magnitude of $J_{\text{inter}}^{\perp,\sigma}$ justifies its omission in the subsequent atomistic spin simulations, where the bilayer can be accurately approximated as two effectively decoupled magnetic subsystems. Therefore, we have ignored them in our atomistic spin simulations for simplicity.

\emph{}

\section*{ACKNOWLEDGMENTS} 
This study has been supported through ANR Grant No. ANR-22-CE24-0019. This work is supported by France 2030 government investment plan managed by the French National Research Agency under grant reference PEPR SPIN – [SPINTHEORY] ANR-22-EXSP-0009. We gratefully acknowledge financial support from the Deutsche Forschungs Gemeinschaft (DFG, German Research Foundation) through SPP2137 “Skyrmionics” (Project No. 462602351). H.S. acknowledges financial support from the Icelandic Research Fund (Grant No. 239435). This work was performed using HPC resources from CALMIP (Grant 2024/2026-[P21008]) and HPC resources available at the Kiel University Computing Centre. M.A.G. thanks V. Kuchkin , P. F. Bessarab, A. Bernand-Mantel, and C. B. Muratov for valuable discussions.

\section*{AUTHOR CONTRIBUTIONS} 
D.Li and S. Heinze conceived and supervised the project. D.Li performed the DFT calculations. M.A.G., D.Li, and S. Haldar performed the GNEB simulations. M.A.G. analyzed soliton localization and calculated lifetimes. D.Li and T.D. worked out the Heisenberg model fit from DFT total energies. M.A.G. and H.S. calculated the topological densities for the honeycomb lattice. M.A.G. and D.Li prepared the figures. M.A.G., S. Heinze, and D.Li wrote the paper, and all authors contributed to the revision of the manuscript.

\emph{}

The authors declare no competing interests.

\bibliography{References}

@misc{fleurv26,
	title = {{Welcome to the FLEUR-project}},
	howpublished = {www.flapw.de (accessed Sept. 1, 2022)},}

@article{Kurz2004,
	title = {\textit{Ab initio} treatment of noncollinear magnets with the full-potential linearized augmented plane wave method},
	author = {Kurz, Ph. and F\"orster, F. and Nordstr\"om, L. and Bihlmayer, G. and Bl\"ugel, S.},
	journal = {Phys. Rev. B},
	volume = {69},
	issue = {2},
	pages = {024415},
	numpages = {15},
	year = {2004},
	month = {Jan},
	publisher = {American Physical Society},
	doi = {10.1103/PhysRevB.69.024415},
}

@article{Chen2025,
  title = {Ferroelectricity-driven metallicity and magnetic skyrmions in the van der {Waals} {${\mathrm{Cr}}_{2}{\mathrm{Ge}}_{2}{\mathrm{Te}}_{6}/{\mathrm{Hf}}_{2}{\mathrm{Ge}}_{2}{\mathrm{Te}}_{6}$} multiferroic heterostructure},
  author = {Chen, Zheng and Hu, Hongliang and Wu, Xiaoping and Zhang, Wenjun and Li, Ping and Song, Changsheng},
  journal = {Phys. Rev. B},
  volume = {111},
  issue = {14},
  pages = {144418},
  numpages = {10},
  year = {2025},
  month = {Apr},
  publisher = {American Physical Society},
  doi = {10.1103/PhysRevB.111.144418}
}

@article{Bo2025,
  title={All-electric control of skyrmion-bimeron transition in van der {Waals} heterostructures},
  author={Bo, Lan and Dai, Songli and Zhang, Xichao and Mochizuki, Masahito and Xu, Xiaohong and Tian, Zean and Zhou, Yan},
  journal={Commun. Phys.},
  volume={8},
  number={1},
  pages={325},
  year={2025},
doi = {10.1038/s42005-025-02224-9},
  publisher={Nature Publishing Group UK London}
}

@article{Zhang2020a,
  title = {Static and dynamic properties of bimerons in a frustrated ferromagnetic monolayer},
  author = {Zhang, Xichao and Xia, Jing and Shen, Laichuan and Ezawa, Motohiko and Tretiakov, Oleg A. and Zhao, Guoping and Liu, Xiaoxi and Zhou, Yan},
  journal = {Phys. Rev. B},
  volume = {101},
  issue = {14},
  pages = {144435},
  numpages = {14},
  year = {2020},
  month = {Apr},
  publisher = {American Physical Society},
  doi = {10.1103/PhysRevB.101.144435}
}

@article{burch2018magnetism,
  title={Magnetism in two-dimensional van der {Waals} materials},
  author={Burch, Kenneth S and Mandrus, David and Park, Je-Geun},
  journal={Nature},
  volume={563},
  number={7729},
  pages={47--52},
  year={2018},
doi = {10.1038/s41586-018-0631-z},
  publisher={Nature Publishing Group UK London}
}

@article{sampaio2013nucleation,
	title={Nucleation, stability and current-induced motion of isolated magnetic skyrmions in nanostructures},
	author={Sampaio, Jo{\~a}o and Cros, Vincent and Rohart, Stanislas and Thiaville, Andr{\'e} and Fert, Albert},
	journal={Nat. Nanotechnol.},
	volume={8},
	number={11},
	pages={839--844},
	year={2013},
doi = {10.1038/nnano.2013.210},
	publisher={Nature Publishing Group UK London}
}

@article{hanneken2015electrical,
	title={Electrical detection of magnetic skyrmions by tunnelling non-collinear magnetoresistance},
	author={Hanneken, Christian and Otte, Fabian and Kubetzka, Andr{\'e} and Dup{\'e}, Bertrand and Romming, Niklas and Von Bergmann, Kirsten and Wiesendanger, Roland and Heinze, Stefan},
	journal={Nat. Nanotechnol.},
	volume={10},
	number={12},
	pages={1039--1042},
	year={2015},
doi = {10.1038/nnano.2015.218},
	publisher={Nature Publishing Group}
}

@article{Perini2019,
	title = {Electrical Detection of Domain Walls and Skyrmions in {Co} Films Using Noncollinear Magnetoresistance},
	author = {Perini, Marco and Meyer, Sebastian and Kubetzka, Andr\'e and Wiesendanger, Roland and Heinze, Stefan and von Bergmann, Kirsten},
	journal = {Phys. Rev. Lett.},
	volume = {123},
	issue = {23},
	pages = {237205},
	numpages = {5},
	year = {2019},
	month = {Dec},
	publisher = {American Physical Society},
	doi = {10.1103/PhysRevLett.123.237205},
}

@article{maccariello2018electrical,
	title={Electrical detection of single magnetic skyrmions in metallic multilayers at room temperature},
	author={Maccariello, Davide and Legrand, William and Reyren, Nicolas and Garcia, Karin and Bouzehouane, Karim and Collin, Sophie and Cros, Vincent and Fert, Albert},
	journal={Nat. Nanotechnol.},
	volume={13},
	number={3},
	pages={233--237},
	year={2018},
doi = {10.1038/s41565-017-0044-4},
	publisher={Nature Publishing Group UK London}
}

@article{buttner2021observation,
	title={Observation of fluctuation-mediated picosecond nucleation of a topological phase},
	author={Buttner, Felix and Pfau, Bastian and Bottcher, Marie and Schneider, Michael and Mercurio, Giuseppe and Gunther, Christian M and Hessing, Piet and Klose, Christopher and Wittmann, Angela and Gerlinger, Kathinka and others},
	journal={Nat. Mater.},
	volume={20},
	number={1},
	pages={30--37},
	year={2021},
doi = {10.1038/s41563-020-00807-1},
	publisher={Nature Publishing Group UK London}
}

@article{dabrowski2022all,
	title={All-optical control of spin in a {2D} van der {W}aals magnet},
	author={Dabrowski, Maciej and Guo, Shi and Strungaru, Mara and Keatley, Paul S and Withers, Freddie and Santos, Elton JG and Hicken, Robert J},
	journal={Nat. Commun.},
	volume={13},
	number={1},
	pages={5976},
	year={2022},
doi = {10.1038/s41467-022-33343-4},
	publisher={Nature Publishing Group UK London}
}

@article{khela2023laser,
	title={Laser-induced topological spin switching in a {2D} van der {W}aals magnet},
	author={Khela, Maya and Dabrowski, Maciej and Khan, Safe and Keatley, Paul S and Verzhbitskiy, Ivan and Eda, Goki and Hicken, Robert J and Kurebayashi, Hidekazu and Santos, Elton JG},
	journal={Nat. Commun.},
	volume={14},
	number={1},
	pages={1378},
	year={2023},
doi = {10.1038/s41467-023-37082-y},
	publisher={Nature Publishing Group UK London}
}

@article{fert2013,
    title = {Skyrmions on the track},
	author={Fert, Albert and Cros, Vincent and Sampaio, Joao},
	journal={Nat. Nanotechnol.},
	volume={8},
	pages={152--156},
	year={2013},
	publisher={Nature Publishing Group},
	doi = {10.1038/nnano.2013.29},
}

@article{Haldar2018,
	title = {First-principles prediction of sub-10-nm skyrmions in {Pd/Fe} bilayers on {Rh(111)}},
	author = {Haldar, Soumyajyoti and von Malottki, Stephan and Meyer, Sebastian and Bessarab, Pavel F. and Heinze, Stefan},
	journal = {Phys. Rev. B},
	volume = {98},
	issue = {6},
	pages = {060413(R)},
	numpages = {6},
	year = {2018},
	month = {Aug},
	publisher = {American Physical Society},
	doi = {10.1103/PhysRevB.98.060413},
}

@article{Malottki2019,
	title = {Skyrmion lifetime in ultrathin films},
	author = {von Malottki, Stephan and Bessarab, Pavel F. and Haldar, Soumyajyoti and Delin, Anna and Heinze, Stefan},
	journal = {Phys. Rev. B},
	volume = {99},
	issue = {6},
	pages = {060409(R)},
	numpages = {5},
	year = {2019},
	month = {Feb},
	publisher = {American Physical Society},
	doi = {10.1103/PhysRevB.99.060409},
}

@article{muckel2021experimental,
	title={Experimental identification of two distinct skyrmion collapse mechanisms},
	author={Muckel, Florian and von Malottki, Stephan and Holl, Christian and Pestka, Benjamin and Pratzer, Marco and Bessarab, Pavel F and Heinze, Stefan and Morgenstern, Markus},
	journal={Nat. Phys.},
	volume={17},
	number={3},
	pages={395--402},
	year={2021},
doi = {10.1038/s41567-020-01101-2},
	publisher={Nature Publishing Group}
}

@article{bessarab2018lifetime,
  title={Lifetime of racetrack skyrmions},
  author={Bessarab, Pavel F and M{\"u}ller, Gideon P and Lobanov, Igor S and Rybakov, Filipp N and Kiselev, Nikolai S and J{\'o}nsson, Hannes and Uzdin, Valery M and Bl{\"u}gel, Stefan and Bergqvist, Lars and Delin, Anna},
  journal={Sci. Rep.},
  volume={8},
  number={},
  pages={3433},
  year={2018},
doi = {10.1038/s41598-018-21623-3},
  publisher={Nature Publishing Group}
}

@article{Desplat2020,
  title = {Path sampling for lifetimes of metastable magnetic skyrmions and direct comparison with Kramers' method},
  author = {Desplat, L. and Vogler, C. and Kim, J.-V. and Stamps, R. L. and Suess, D.},
  journal = {Phys. Rev. B},
  volume = {101},
  issue = {6},
  pages = {060403},
  numpages = {5},
  year = {2020},
  month = {Feb},
  publisher = {American Physical Society},
  doi = {10.1103/PhysRevB.101.060403}
}

@article{meyer2019isolated,
	title={Isolated zero field sub-10 nm skyrmions in ultrathin {Co} films},
	author={Meyer, Sebastian and Perini, Marco and von Malottki, Stephan and Kubetzka, Andr{\'e} and Wiesendanger, Roland and von Bergmann, Kirsten and Heinze, Stefan},
	journal={Nat. Commun.},
	volume={10},
	number={1},
	pages={3823},
	year={2019},
doi = {10.1038/s41467-019-11831-4},
	publisher={Nature Publishing Group}
}

@article{Zhuang2016,
	title = {Strong anisotropy and magnetostriction in the two-dimensional {Stoner} ferromagnet {Fe$_3$GeTe$_2$}},
	author = {Zhuang, Houlong L. and Kent, P. R. C. and Hennig, Richard G.},
	journal = {Phys. Rev. B},
	volume = {93},
	issue = {13},
	pages = {134407},
	numpages = {7},
	year = {2016},
	month = {Apr},
	publisher = {American Physical Society},
	doi = {10.1103/PhysRevB.93.134407},
}

@article{gong2017discovery,
	title={Discovery of intrinsic ferromagnetism in two-dimensional van der {W}aals crystals},
	author={Gong, Cheng and Li, Lin and Li, Zhenglu and Ji, Huiwen and Stern, Alex and Xia, Yang and Cao, Ting and Bao, Wei and Wang, Chenzhe and Wang, Yuan and others},
	journal={Nature},
	volume={546},
	pages={265--269},
	year={2017},
	publisher={Nature Publishing Group},
	doi = {10.1038/nature22060},
}

@article{dupe2014tailoring,
	title={Tailoring magnetic skyrmions in ultra-thin transition metal films},
	author={Dup{\'e}, Bertrand and Hoffmann, Markus and Paillard, Charles and Heinze, Stefan},
	journal={Nat. Commun.},
	volume={5},
	number={1},
	pages={4030},
	year={2014},
doi = {10.1038/ncomms5030},
	publisher={Nature Publishing Group}
}

@article{hoffmann2017antiskyrmions,
  title={Antiskyrmions stabilized at interfaces by anisotropic {Dzyaloshinskii-Moriya interactions}},
  author={Hoffmann, Markus and Zimmermann, Bernd and M{\"u}ller, Gideon P and Sch{\"u}rhoff, Daniel and Kiselev, Nikolai S and Melcher, Christof and Bl{\"u}gel, Stefan},
  journal={Nat. Commun.},
  volume={8},
  number={1},
  pages={308},
  year={2017},
doi = {10.1038/s41467-017-00313-0},
  publisher={Nature Publishing Group UK London}
}

@article{grimme2010consistent,
	title={A consistent and accurate ab initio parametrization of density functional dispersion correction {(DFT-D)} for the 94 elements {H-Pu}},
	author={Grimme, Stefan and Antony, Jens and Ehrlich, Stephan and Krieg, Helge},
	journal={J. Chem. Phys.},
	volume={132},
	number={15},
	pages={154104},
	year={2010},
doi = {10.1063/1.3382344},
	publisher={American Institute of Physics}
}

@article{wu2022van,
	author = {Wu, Yingying and Francisco, Brian and Chen, Zhijie and Wang, Wei and Zhang, Yu and Wan, Caihua and Han, Xiufeng and Chi, Hang and Hou, Yasen and Lodesani, Alessandro and Yin, Gen and Liu, Kai and Cui, Yong-tao and Wang, Kang L. and Moodera, Jagadeesh S.},
	title = {A Van der {W}aals Interface Hosting Two Groups of Magnetic Skyrmions},
	journal = {Adv. Mater.},
	volume = {34},
	number = {16},
	pages = {2110583},
	doi = {10.1002/adma.202110583},
	year = {2022}
}

@article{bessarab2015method,
	title={Method for finding mechanism and activation energy of magnetic transitions, applied to skyrmion and antivortex annihilation},
	author={Bessarab, Pavel F and Uzdin, Valery M and J{\'o}nsson, Hannes},
	journal={Comput. Phys. Commun.},
	volume={196},
	pages={335--347},
	year={2015},
doi = {10.1016/j.cpc.2015.07.001},
	publisher={Elsevier}
}

@article{romming2013writing,
	title={Writing and deleting single magnetic skyrmions},
	author={Romming, Niklas and Hanneken, Christian and Menzel, Matthias and Bickel, Jessica E and Wolter, Boris and von Bergmann, Kirsten and Kubetzka, Andr{\'e} and Wiesendanger, Roland},
	journal={Science},
	volume={341},
	number={6146},
	pages={636--639},
	year={2013},
doi = {10.1126/science.124057},
	publisher={American Association for the Advancement of Science}
}

@article{Dongzhe2022_fgt,
	author = {Li, D. and Haldar, S. and Heinze, S.},
	title = {Strain-Driven Zero-Field Near-10 nm Skyrmions in Two-Dimensional van der {W}aals Heterostructures},
	journal = {Nano Lett.},
	volume = {22},
	number = {18},
	pages = {7706-7713},
	year = {2022},
	doi = {10.1021/acs.nanolett.2c03287},
}

@article{Dongzhe2023,
author = {Li, Dongzhe and Haldar, Soumyajyoti and Heinze, Stefan},
title = {Proposal for All-Electrical Skyrmion Detection in van der {W}aals Tunnel Junctions},
journal = {Nano Lett.},
volume = {24},
number = {8},
pages = {2496-2502},
year = {2024},
doi = {10.1021/acs.nanolett.3c04238},
}

@article{sun2020controlling,
	title={Controlling bimerons as skyrmion analogues by ferroelectric polarization in {2D} van der {W}aals multiferroic heterostructures},
	author={Sun, Wei and Wang, Wenxuan and Li, Hang and Zhang, Guangbiao and Chen, Dong and Wang, Jianli and Cheng, Zhenxiang},
	journal={Nat. Commun.},
	volume={11},
	number={1},
	pages={5930},
	year={2020},
doi = {10.1038/s41467-020-19779-6},
	publisher={Nature Publishing Group}
}

@article{ohara2022reversible,
	title={Reversible Transformation between Isolated Skyrmions and Bimerons},
	author={Ohara, Kentaro and Zhang, Xichao and Chen, Yinling and Kato, Satoshi and Xia, Jing and Ezawa, Motohiko and Tretiakov, Oleg A and Hou, Zhipeng and Zhou, Yan and Zhao, Guoping and others},
	journal={Nano Lett.},
	volume={22},
	number={21},
	pages={8559--8566},
	year={2022},
doi = {10.1021/acs.nanolett.2c03106},
	publisher={ACS Publications}
}

@article{Dongzhe_prb2023,
	title = {Tuning the magnetic interactions in van der {W}aals {${\mathrm{Fe}}_{3}{\mathrm{GeTe}}_{2}$} heterostructures: A comparative study of \textit{ab initio} methods},
	author = {Li, Dongzhe and Haldar, Soumyajyoti and Drevelow, Tim and Heinze, Stefan},
	journal = {Phys. Rev. B},
	volume = {107},
	issue = {10},
	pages = {104428},
	numpages = {15},
	year = {2023},
	month = {Mar},
	publisher = {American Physical Society},
	doi = {10.1103/PhysRevB.107.104428},
}

@article{Dongzhe_PRB2024,
  title = {Prediction of stable nanoscale skyrmions in monolayer {Fe$_5$GeTe$_2$}},
  author = {Li, Dongzhe and Haldar, Soumyajyoti and Kollwitz, Leo and Schrautzer, Hendrik and Goerzen, Moritz A. and Heinze, Stefan},
  journal = {Phys. Rev. B},
  volume = {109},
  issue = {22},
  pages = {L220404},
  numpages = {7},
  year = {2024},
  month = {Jun},
  publisher = {American Physical Society},
  doi = {10.1103/PhysRevB.109.L220404},
}

@article{hagemeister2015stability,
  title={Stability of single skyrmionic bits},
  author={Hagemeister, J and Romming, N and Von Bergmann, K and Vedmedenko, EY and Wiesendanger, R},
  journal={Nat. Commun.},
  volume={6},
  pages={8455},
  year={2015},
doi = {10.1038/ncomms9455},
  publisher={Nature Publishing Group}
}

@article{vosko1980accurate,
	title={Accurate spin-dependent electron liquid correlation energies for local spin density calculations: a critical analysis},
	author={Vosko, Seymour H and Wilk, Leslie and Nusair, Marwan},
	journal={Can. J. Phys.},
	volume={58},
	number={8},
	pages={1200--1211},
	year={1980},
doi = {10.1139/p80-159},
	publisher={NRC Research Press Ottawa, Canada}
}

@article{chen2024all,
  title={All-electrical skyrmionic magnetic tunnel junction},
  author={Chen, Shaohai and Lourembam, James and Ho, Pin and Toh, Alexander KJ and Huang, Jifei and Chen, Xiaoye and Tan, Hang Khume and Yap, Sherry LK and Lim, Royston JJ and Tan, Hui Ru and others},
  journal={Nature},
  volume={627},
  number={8004},
  pages={522--527},
  year={2024},
doi = {10.1038/s41586-024-07131-7},
  publisher={Nature Publishing Group}
}

@article{gobel2021beyond,
	title={Beyond skyrmions: Review and perspectives of alternative magnetic quasiparticles},
	author={G{\"o}bel, B{\"o}rge and Mertig, Ingrid and Tretiakov, Oleg A},
	journal={Phys. Rep.},
	volume={895},
	pages={1--28},
	year={2021},
doi = {10.1016/j.physrep.2020.10.001},
	publisher={Elsevier}
}

@article{fert2017magnetic,
	title={Magnetic skyrmions: advances in physics and potential applications},
	author={Fert, Albert and Reyren, Nicolas and Cros, Vincent},
	journal={Nat. Rev. Mater.},
	volume={2},
	number={7},
	pages={1--15},
	year={2017},
doi = {10.1038/natrevmats.2017.31},
	publisher={Nature Publishing Group}
}

@article{deng2018gate,
	title={Gate-tunable room-temperature ferromagnetism in two-dimensional {Fe$_3$GeTe$_2$}},
	author={Deng, Yujun and Yu, Yijun and Song, Yichen and Zhang, Jingzhao and Wang, Nai Zhou and Sun, Zeyuan and Yi, Yangfan and Wu, Yi Zheng and Wu, Shiwei and Zhu, Junyi and others},
	journal={Nature},
	volume={563},
	number={7729},
	pages={94--99},
	year={2018},
doi = {10.1038/s41586-018-0626-9},
	publisher={Nature Publishing Group}
}

@article{Heide2009,
	author = {Heide, M. and Bihlmayer, G. and Bl{\"{u}}gel, S.},
	journal = {Phys. B},
	number = {18},
	pages = {2678--2683},
	title = {{Describing Dzyaloshinskii-Moriya spirals from first principles}},
	volume = {404},
	year = {2009},
	doi = {10.1016/j.physb.2009.06.070}
}

@article{kuchkin2020magnetic,
	title={Magnetic skyrmions, chiral kinks, and holomorphic functions},
	author={Kuchkin, Vladyslav M and Barton-Singer, Bruno and Rybakov, Filipp N and Bl{\"u}gel, Stefan and Schroers, Bernd J and Kiselev, Nikolai S},
	journal={Phys. Rev. B},
	volume={102},
	number={14},
	pages={144422},
	year={2020},
doi = {10.1103/PhysRevB.102.144422},
	publisher={APS}
}

@article{barton2020magnetic,
	title={Magnetic skyrmions at critical coupling},
	author={Barton-Singer, Bruno and Ross, Calum and Schroers, Bernd J},
	journal={Commun. Math. Phys.},
	volume={375},
	number={3},
	pages={2259--2280},
	year={2020},
doi = {10.1007/s00220-019-03676-1},
	publisher={Springer}
}

@article{schroers2019gauged,
	title={Gauged sigma models and magnetic skyrmions},
	author={Schroers, Bernd},
	journal={SciPost Physics},
	volume={7},
	number={3},
	pages={030},
doi = {10.21468/SciPostPhys.7.3.030},
	year={2019}
}

@article{zarzuela2020stability,
  title={Stability and dynamics of in-plane skyrmions in collinear ferromagnets},
  author={Zarzuela, Ricardo and Bharadwaj, Venkata Krishna and Kim, Kyoung-Whan and Sinova, Jairo and Everschor-Sitte, Karin},
  journal={Phys. Rev. B},
  volume={101},
  number={5},
  pages={054405},
  year={2020},
doi = {10.1103/PhysRevB.101.054405},
  publisher={APS}
}

@article{kharkov2017bound,
  title={Bound states of skyrmions and merons near the {Lifshitz} point},
  author={Kharkov, YA and Sushkov, OP and Mostovoy, M},
  journal={Phys. Rev. Lett.},
  volume={119},
  number={20},
  pages={207201},
  year={2017},
doi = {10.1103/PhysRevLett.119.207201},
  publisher={APS}
}

@article{gobel2019magnetic,
  title={Magnetic bimerons as skyrmion analogues in in-plane magnets},
  author={G{\"o}bel, B{\"o}rge and Mook, Alexander and Henk, J{\"u}rgen and Mertig, Ingrid and Tretiakov, Oleg A},
  journal={Phys. Rev. B},
  volume={99},
  number={6},
  pages={060407},
  year={2019},
doi = {10.1103/PhysRevB.99.060407},
  publisher={APS}
}

@article{bocdanov1994properties,
  title={The properties of isolated magnetic vortices},
  author={Bocdanov, A and Hubert, A},
  journal={Phys. Status Solidi (b)},
  volume={186},
  number={2},
  pages={527--543},
  year={1994},
doi = {10.1002/pssb.2221860223},
  publisher={Wiley Online Library}
}

@article{varentcova2020toward,
  title={Toward room-temperature nanoscale skyrmions in ultrathin films},
  author={Varentcova, Anastasiia S and von Malottki, Stephan and Potkina, Maria N and Kwiatkowski, Grzegorz and Heinze, Stefan and Bessarab, Pavel F},
  journal={npj Comput. Mater.},
  volume={6},
  number={},
  pages={193},
  year={2020},
doi = {10.1038/s41524-020-00453-w},
  publisher={Nature Publishing Group}
}

@article{nagaosa2013topological,
  title={Topological properties and dynamics of magnetic skyrmions},
  author={Nagaosa, Naoto and Tokura, Yoshinori},
  journal={Nat. Nanotechnol.},
  volume={8},
  number={12},
  pages={899--911},
  year={2013},
doi = {10.1038/nnano.2013.243},
  publisher={Nature Publishing Group}
}

@article{goerzen2023lifetime,
  title={Lifetime of coexisting sub-10 nm zero-field skyrmions and antiskyrmions},
  author={Goerzen, Moritz A and von Malottki, Stephan and Meyer, Sebastian and Bessarab, Pavel F and Heinze, Stefan},
  journal={npj Quantum Mater.},
  volume={8},
  pages={54},
 doi={10.1038/s41535-023-00586-3},
  year={2023}
}

@article{kuchkin2020turning,
  title = {Turning a chiral skyrmion inside out},
  author = {Kuchkin, Vladyslav M. and Kiselev, Nikolai S.},
  journal = {Phys. Rev. B},
  volume = {101},
  issue = {6},
  pages = {064408},
  numpages = {10},
  year = {2020},
  month = {Feb},
  publisher = {American Physical Society},
  doi = {10.1103/PhysRevB.101.064408},
}

@article{berg1981,
  title={Definition and statistical distributions of a topological number in the lattice {O(3)} $\sigma$-model},
  author={Berg, B and L{\"u}scher, Martin},
  journal={Nucl. Phys. B},
  volume={190},
  number={2},
  pages={412--424},
  year={1981},
doi = {10.1016/0550-3213(81)90568-X},
  publisher={Elsevier}
}

@article{Shen2020,
  title = {Current-Induced Dynamics and Chaos of Antiferromagnetic Bimerons},
  author = {Shen, Laichuan and Xia, Jing and Zhang, Xichao and Ezawa, Motohiko and Tretiakov, Oleg A. and Liu, Xiaoxi and Zhao, Guoping and Zhou, Yan},
  journal = {Phys. Rev. Lett.},
  volume = {124},
  issue = {3},
  pages = {037202},
  numpages = {6},
  year = {2020},
  month = {Jan},
  publisher = {American Physical Society},
  doi = {10.1103/PhysRevLett.124.037202}
}

@article{jani2021antiferromagnetic,
  title={Antiferromagnetic half-skyrmions and bimerons at room temperature},
  author={Jani, Hariom and Lin, Jheng-Cyuan and Chen, Jiahao and Harrison, Jack and Maccherozzi, Francesco and Schad, Jonathon and Prakash, Saurav and Eom, Chang-Beom and Ariando, Ariando and Venkatesan, Thirumalai and others},
  journal={Nature},
  volume={590},
  number={7844},
  pages={74--79},
  year={2021},
doi = {10.1038/s41586-021-03219-6},
  publisher={Nature Publishing Group UK London}
}

@article{bessarab2012harmonic,
  title={Harmonic transition-state theory of thermal spin transitions},
  author={Bessarab, Pavel F and Uzdin, Valery M and J{\'o}nsson, Hannes},
  journal={Phys. Rev. B},
  volume={85},
  number={18},
  pages={184409},
  year={2012},
doi = {10.1103/PhysRevB.85.184409},
  publisher={APS}
}

@article{goerzen2022atomistic,
  title={Atomistic spin simulations of electric-field-assisted nucleation and annihilation of magnetic skyrmions in {Pd/Fe/Ir (111)}},
  author={Goerzen, Moritz A and von Malottki, Stephan and Kwiatkowski, Grzegorz J and Bessarab, Pavel F and Heinze, Stefan},
  journal={Phys. Rev. B},
  volume={105},
  number={21},
  pages={214435},
  year={2022},
doi = {10.1103/PhysRevB.105.214435},
  publisher={APS}
}

@article{rohart2013skyrmion,
  title={Skyrmion confinement in ultrathin film nanostructures in the presence of {Dzyaloshinskii-Moriya} interaction},
  author={Rohart, S and Thiaville, A},
  journal={Phys. Rev. B},
  volume={88},
  number={18},
  pages={184422},
  year={2013},
doi = {10.1103/PhysRevB.88.184422},
  publisher={APS}
}

@article{polyakov22metastable,
  title={Metastable states of two-dimensional isotropic ferromagnets 1975},
  author={Polyakov, AM and Belavin, AA},
  journal={Sov. Phys. JETP Lett},
  volume={22},
  pages={245}
}

@article{pinna2018,
  title={Skyrmion gas manipulation for probabilistic computing},
  author={Pinna, Daniele and Abreu Araujo, Flavio and Kim, J-V and Cros, Vincent and Querlioz, Damien and Bessiere, Pierre and Droulez, Jacques and Grollier, Julie},
  journal={Phys. Rev. Appl.},
  volume={9},
  number={6},
  pages={064018},
  year={2018},
doi = {10.1103/PhysRevApplied.9.064018},
  publisher={APS}
}

@article{song2020,
  title={Skyrmion-based artificial synapses for neuromorphic computing},
  author={Song, Kyung Mee and Jeong, Jae-Seung and Pan, Biao and Zhang, Xichao and Xia, Jing and Cha, Sunkyung and Park, Tae-Eon and Kim, Kwangsu and Finizio, Simone and Raabe, J{\"o}rg and others},
  journal={Nat. Electron.},
  volume={3},
  number={3},
  pages={148--155},
  year={2020},
doi = {10.1038/s41928-020-0385-0},
  publisher={Nature Publishing Group UK London}
}

@article{wang2025magnetoelectric,
  title={Magnetoelectric Bimeron in {2D} Hexagonal Lattice},
  author={Wang, Xiudong and He, Zhonglin and Dai, Ying and Huang, Baibiao and Ma, Yandong},
  journal={Adv. Funct. Mater},
  pages={e10581},
  year={2025},
doi = {10.1002/adfm.202510581},
  publisher={Wiley Online Library}
}

@article{skyquibit_2021,
  title = {Skyrmion Qubits: A New Class of Quantum Logic Elements Based on Nanoscale Magnetization},
  author = {Psaroudaki, Christina and Panagopoulos, Christos},
  journal = {Phys. Rev. Lett.},
  volume = {127},
  issue = {6},
  pages = {067201},
  numpages = {6},
  year = {2021},
  month = {Aug},
  publisher = {American Physical Society},
  doi = {10.1103/PhysRevLett.127.067201},
}

@article{bramwell1994magnetization,
  title = {Magnetization: A characteristic of the {Kosterlitz-Thouless-Berezinskii} transition},
  author = {Bramwell, S. T. and Holdsworth, P. C. W.},
  journal = {Phys. Rev. B},
  volume = {49},
  issue = {13},
  pages = {8811--8814},
  numpages = {0},
  year = {1994},
  month = {Apr},
  publisher = {American Physical Society},
  doi = {10.1103/PhysRevB.49.8811},
}

@article{deng2025,
  title={The conformal limit for bimerons in easy-plane chiral magnets},
  author={Deng, Bin and Ignat, Radu and Lamy, Xavier},
  journal={arXiv preprint arXiv:2506.11955},
url= {https://arxiv.org/abs/2506.11955},
  year={2025}
}

@article{ivanov2007quantum,
  title={Quantum effects for the two-dimensional soliton in isotropic ferromagnets},
  author={Ivanov, Boris A and Sheka, Denis D and Kryvonos, Vasyl V and Mertens, Franz G},
  journal={Phys. Rev. B},
  volume={75},
  number={13},
  pages={132401},
  year={2007},
doi = {10.1103/PhysRevB.75.132401},
  publisher={APS}
}

@article{schrautzer2022effects,
  title={Effects of interlayer exchange on collapse mechanisms and stability of magnetic skyrmions},
  author={Schrautzer, Hendrik and von Malottki, Stephan and Bessarab, Pavel F and Heinze, Stefan},
  journal={Phys. Rev. B},
  volume={105},
  number={1},
  pages={014414},
  year={2022},
doi = {10.1103/PhysRevB.105.014414},
  publisher={APS}
}

@article{nagase2021observation,
  title={Observation of domain wall bimerons in chiral magnets},
  author={Nagase, Tomoki and So, Yeong-Gi and Yasui, Hayata and Ishida, Takafumi and Yoshida, Hiroyuki K and Tanaka, Yukio and Saitoh, Koh and Ikarashi, Nobuyuki and Kawaguchi, Yuki and Kuwahara, Makoto and others},
  journal={Nat. Commun.},
  volume={12},
  number={1},
  pages={3490},
  year={2021},
doi = {10.1038/s41467-021-23845-y },
  publisher={Nature Publishing Group UK London}
}

@article{lin2016ginzburg,
  title={Ginzburg-{L}andau theory for skyrmions in inversion-symmetric magnets with competing interactions},
  author={Lin, Shi-Zeng and Hayami, Satoru},
  journal={Phys. Rev. B},
  volume={93},
  number={6},
  pages={064430},
  year={2016},
doi = {10.1103/PhysRevB.93.064430},
  publisher={APS}
}

@article{potkina2020skyrmions,
  title={Skyrmions in antiferromagnets: Thermal stability and the effect of external field and impurities},
  author={Potkina, Maria N and Lobanov, Igor S and J{\'o}nsson, Hannes and Uzdin, Valery M},
  journal={J. Appl. Phys.},
  volume={127},
  number={21},
  year={2020},
doi = {10.1063/5.0009559
},
  publisher={AIP Publishing}
}

@article{yu2024spontaneous,
  title={Spontaneous Vortex-Antivortex Pairs and Their Topological Transitions in a Chiral-Lattice Magnet},
  author={Yu, Xiuzhen and Kanazawa, Naoya and Zhang, Xichao and Takahashi, Yoshio and Iakoubovskii, Konstantin V and Nakajima, Kiyomi and Tanigaki, Toshiaki and Mochizuki, Masahito and Tokura, Yoshinori},
  journal={Adv. Mater.},
  volume={36},
  number={1},
  pages={2306441},
  year={2024},
  doi = {10.1002/adma.202306441},
  publisher={Wiley Online Library}
}

@article{hohenberg1967existence,
  title = {Existence of Long-Range Order in One and Two Dimensions},
  author = {Hohenberg, P. C.},
  journal = {Phys. Rev.},
  volume = {158},
  issue = {2},
  pages = {383--386},
  numpages = {0},
  year = {1967},
  month = {Jun},
  publisher = {American Physical Society},
  doi = {10.1103/PhysRev.158.383}
}

@article{mermin1966absence,
  title = {Absence of Ferromagnetism or Antiferromagnetism in One- or Two-Dimensional Isotropic Heisenberg Models},
  author = {Mermin, N. D. and Wagner, H.},
  journal = {Phys. Rev. Lett.},
  volume = {17},
  issue = {22},
  pages = {1133--1136},
  numpages = {0},
  year = {1966},
  month = {Nov},
  publisher = {American Physical Society},
  doi = {10.1103/PhysRevLett.17.1133}
}

@article{jiang2017direct,
  title={Direct observation of the skyrmion {Hall} effect},
  author={Jiang, Wanjun and Zhang, Xichao and Yu, Guoqiang and Zhang, Wei and Wang, Xiao and Benjamin Jungfleisch, M and Pearson, John E and Cheng, Xuemei and Heinonen, Olle and Wang, Kang L and others},
  journal={Nat. Phys.},
  volume={13},
  number={2},
  pages={162--169},
  year={2017},
doi = {10.1038/nphys3883},
  publisher={Nature Publishing Group UK London}
}

@article{rybakov2025topological,
  title = {Topological invariants of vortices, merons, skyrmions, and their combinations in continuous and discrete systems},
  author = {Rybakov, Filipp N. and Eriksson, Olle and Kiselev, Nikolai S.},
  journal = {Phys. Rev. B},
  volume = {111},
  issue = {13},
  pages = {134417},
  numpages = {13},
  year = {2025},
  month = {Apr},
  publisher = {American Physical Society},
  doi = {10.1103/PhysRevB.111.134417},
  url = {https://link.aps.org/doi/10.1103/PhysRevB.111.134417}
}

@article{zhu2026light,
  title={Light-induced bimerons in a chiral magnet},
  author={Zhu, Kaixin and Rybakov, Filipp N and Wang, Zhan and Gao, Wenli and Sun, Shuaishuai and Wang, Wentao and Li, Jun and Tian, Huanfang and Eriksson, Olle and Yang, Huaixin and others},
  journal={Nat. Commun.},
  year         = {2026},
  volume       = {17},
  pages        = {3185},
  doi          = {10.1038/s41467-026-71291-5},
  publisher={Nature Publishing Group UK London}
}

@article{Masell2019,
  title = {Universality of annihilation barriers of large magnetic skyrmions in chiral and frustrated magnets},
  author = {Heil, Benjamin and Rosch, Achim and Masell, Jan},
  journal = {Phys. Rev. B},
  volume = {100},
  issue = {13},
  pages = {134424},
  numpages = {14},
  year = {2019},
  month = {Oct},
  publisher = {American Physical Society},
  doi = {10.1103/PhysRevB.100.134424},
  url = {https://link.aps.org/doi/10.1103/PhysRevB.100.134424}
}

@article{zhu2026strongly,
author = {Zhu, Shiwei and Goerzen, Moritz A. and Song, Changsheng and Heinze, Stefan and Li, Dongzhe},
title = {Strongly Enhanced Lifetime of Higher-Order Bimerons and Antibimerons},
journal = {Nano Lett.},
volume = {26},
number = {19},
pages = {6307},
year = {2026},
doi = {10.1021/acs.nanolett.6c00092},
}

@article{schrautzer2026impact,
  title={Impact of higher-order exchange on the lifetime of skyrmions and antiskyrmions},
  author={Schrautzer, Hendrik and Goerzen, Moritz A and Beyer, Bjarne and Haldar, Soumyajyoti and Bessarab, Pavel F and Heinze, Stefan},
  journal={Npj Comput. Mater.},
  year={2026},
  doi={10.1038/s41524-026-02034-9},
  volume = {12},
pages = {123},
  publisher={Nature Publishing Group UK London}
}

@article{kosterlitz1973ordering,
  title={Ordering, metastability and phase transitions in two-dimensional systems},
  author={Kosterlitz, John Michael and Thouless, David James},
  journal={J. Phys. C: Solid State Physics},
  volume={6},
  number={7},
  pages={1181--1203},
  doi={10.1088/0022-3719/6/7/010},
  year={1973}
}

@article{liao2024tunable,
  title = {Tunable meron pair excitations and {Berezinskii-Kosterlitz-Thouless} phase transitions in the monolayer antiferromagnet {$\mathrm{MnPS}{\mathrm{e}}_{3}$}},
  author = {Liao, Guokai and Zhang, Shunhong and Cui, Ping and Zhang, Zhenyu},
  journal = {Phys. Rev. B},
  volume = {109},
  issue = {10},
  pages = {L100403},
  numpages = {9},
  year = {2024},
  month = {Mar},
  publisher = {American Physical Society},
  doi = {10.1103/PhysRevB.109.L100403},
  url = {https://link.aps.org/doi/10.1103/PhysRevB.109.L100403}
}

@article{schwab2025skyrmion,
  title={Skyrmion bags of light in plasmonic moir{\'e} superlattices},
  author={Schwab, Julian and Neuhaus, Alexander and Dreher, Pascal and Tsesses, Shai and Cohen, Kobi and Mangold, Florian and Mantha, Anant and Frank, Bettina and Bartal, Guy and Meyer zu Heringdorf, Frank-J and others},
  journal={Nat. Phys.},
  volume={21},
  number={6},
  pages={988--994},
  year={2025},
  doi={10.1038/s41567-025-02873-1},
  publisher={Nature Publishing Group UK London}
}

\end{document}